\documentstyle[12pt,epsf]{article}

\advance \topmargin by -\headheight
\advance \topmargin by -\headsep

\evensidemargin \oddsidemargin
\def\ie{{\em i.e.}}

\def\ie{\hbox{\it i.e.}}

\def\CC{{\mathchoice
{\rm C\mkern-8mu\vrule height1.45ex depth-.05ex 
width.05em\mkern9mu\kern-.05em}
{\rm C\mkern-8mu\vrule height1.45ex depth-.05ex 
width.05em\mkern9mu\kern-.05em}
{\rm C\mkern-8mu\vrule height1ex depth-.07ex 
width.035em\mkern9mu\kern-.035em}
{\rm C\mkern-8mu\vrule height.65ex depth-.1ex 
width.025em\mkern8mu\kern-.025em}}}

\def\RR{{\rm I\kern-1.6pt {\rm R}}}

\def\ZZ{{\rm Z}\kern-3.8pt {\rm Z} \kern2pt}

\def\np{Nucl. Phys.}
\def\pl{Phys. Lett.}

\def\pr{Phys. Rev.}

\def\cmp{Comm. Math. Phys.}
\def\jmp{J. Math. Phys.}
\def\ijmp{Int. J. Mod. Phys.}

\def\adm{Adv. in Math.}

\def\jept{Sov. Phys. JEPT}
\def\tmp{Theor. Math. Phys.}

\newcommand{\beq}{\begin{equation}}
\newcommand{\eeq}{\end{equation}}
\newcommand{\rc}{\nonumber\\}
\newcommand{\bear}{\begin{eqnarray}}
\newcommand{\eear}{\end{eqnarray}}

\newfont{\namefont}{cmr10}
\newfont{\addfont}{cmti7 scaled 1440}
\newfont{\boldmathfont}{cmbx10}
\newfont{\headfontb}{cmbx10 scaled 1728}
\renewcommand{\theequation}{{\rm\thesection.\arabic{equation}}}
\begin{document}
\begin{titlepage}

\begin{center} \Large \bf Graded Parafermions

\end{center}

\vskip 0.3truein
\begin{center} 
J. M. Camino
\footnote{e-mail:camino@gaes.usc.es}, 
A.V. Ramallo
\footnote{e-mail:alfonso@gaes.usc.es}
and 
J. M. S\'anchez de Santos 
\footnote{e-mail:santos@gaes.usc.es}

\vspace{0.3in}

Departamento de F\'\i sica de
Part\'\i culas, \\ Universidad de Santiago\\
E-15706 Santiago de Compostela, Spain. 
\vspace{0.3in}

\end{center}
\vskip 1truein

\begin{center}
\bf ABSTRACT
\end{center} 

A graded generalization of the $Z_k$ parafermionic current
algebra is constructed. This symmetry is realized in the 
${\rm osp}(1\vert 2)/U(1)$ coset conformal field theory.  
The structure of the parafermionic highest-weight modules is
analyzed and the dimensions of the fields of the theory are
determined. A free field realization of the graded parafermionic
system is obtained and the structure constants of the current
algebra are found. Although the  theory is not unitary, it
presents good reducibility properties.

\vskip6.5truecm
\leftline{US-FT-7/98 \hfill May 1998}
\leftline{hep-th/9805160}
\smallskip
\end{titlepage}
\setcounter{footnote}{0}

\setcounter{equation}{0}
\section{Introduction}
\medskip

There is no doubt of the importance of symmetries in quantum
field theory. Indeed, the identification of the various
invariances of a system is a fundamental step in the
understanding of its dynamics. In two-dimensional Conformal Field
Theory (CFT) the symmetries are generated by primary operators
which, together with the energy-momentum tensor, generate the
chiral algebra of the model \cite{CFT}. The Hilbert space of the
theory can be described by means of the representation theory of
the chiral algebra, which determines the dimensions of the fields,
the selection rules and, generally speaking, the correlation
functions of the theory.

The parafermionic symmetry is generated by non-local 
currents which obey fractional statistics \cite{para}. In the CFT
context, the parafermions were introduced in ref. \cite{ZF}, as a
generalization of the Ising model. These CFTs possess a $Z_k$
symmetry and describe self-dual critical points in statistical
mechanics. An important observation made in ref. \cite{ZF}  is
that the $Z_k$ parafermionic system can be 
regarded as the $su(2)/U(1)$
coset model \cite{GKO}, $k$ being the level of the affine $su(2)$
algebra. This fact can be used to construct generalized
parafermions based on arbitrary Lie algebras \cite{Gepner}.  Some
other aspects of the parafermionic symmetry have been studied in
refs. \cite{ZFdos}-\cite{Ahn}.

In this paper we construct a graded generalization of the $Z_k$
parafermions. In our system, apart from the currents with integer
charges introduced in ref. \cite{ZF} (which we shall consider as
Grassmann even), we shall have additional currents with
half-integer charges and odd Grassmann parity. These Grassmann
parities are a crucial ingredient in our construction. Actually,
depending on their relative Grassmann character, an extra sign
can appear in the generalized exchange relation obeyed by the
currents. We will see that, making use of very general arguments,
one can find the dimensions of the currents and the general form
of the parafermionic algebra. After examining in detail these
results, it is not hard to conclude that the graded parafermionic
symmetry is related to the ${\rm osp}(1\vert 2)$ affine Lie
superalgebra \cite{Pais, Review}. This relation is similar to the
one that exists between the $Z_k$ parafermions and the $su(2)$
current algebra,
\ie\ our graded parafermionic model can be obtained by modding out
in the 
${\rm osp}(1\vert 2)$ Kac-Moody symmetry  the dependence on the
Cartan subalgebra. This process generates the 
${\rm osp}(1\vert 2)/U(1)$ coset CFT, from which an explicit
realization of the parafermionic algebra can be easily obtained.

The organization of this paper is the following. In section 2 we
shall characterize in general terms the graded parafermionic
symmetry and we will identify it with the one realized in the 
${\rm osp}(1\vert 2)/U(1)$ coset theory. From this identification
we will be able to determine the central charge of the model. In
section 3 we shall analyze the structure of the Hilbert space of
the theory. We will obtain in this section the generalized
(anti)commutation relations  satisfied by the mode
operators of the currents in the different charge sectors of the
theory. This study will serve us to determine the conformal
dimensions of the fields associated to the highest-weight
parafermionic modules.

In section 4 we will present a free field realization of the
parafermionic currents. This realization is based on the one
proposed in ref. \cite{ber} for the ${\rm osp}(1\vert 2)$ current
algebra and worked out in detail in ref. \cite{osp} (see also
refs. \cite{ospdos, rama}). In particular, this free field
representation and the results of ref.
\cite{osp} will be used to compute some correlation functions   of
the parafermionic fields. The structure constants of the
parafermionic current algebra are determined in section 5. We will
combine in this analysis explicit calculations performed by using
the free field realization of the currents and consistency
conditions imposed by the associativity of the operator product
algebra. Some details of these calculations are given in appendix
A. The final expression for the structure constants is remarkably
simple and  generalizes the one found in ref. \cite{ZF} for  the 
$Z_k$ parafermions. Finally, in section 6 we summarize our
results and explore some possible directions of future work.

\setcounter{equation}{0}
\section{The Graded Parafermionic Algebra}
\medskip
Let us consider a two dimensional Conformal Field Theory and
let us concentrate on the holomorphic sector of the theory. We
shall say that the theory is endowed with a parafermionic
algebra if there exists a set of fractional spin currents
$\psi_l$, labelled by a charge quantum number $l$. These
currents are primary fields that obey a generalized fractional
statistics, \ie\ when they are exchanged in a radially ordered
product the latter is multiplied by a root of unity. This
behaviour is a generalization of that of ordinary bosons and
fermions. Moreover, under multiplication the parafermionic
currents behave additively with respect to their charges, \ie\
by multiplying $\psi_l$ and $\psi_{l'}$ the current
$\psi_{l+l'}$ is generated. 

In what follows we shall study a system in which the
parafermionic currents are graded  according to a $\ZZ_2$
Grassmann quantum number. In order to determine which currents
are Grassmann even and which ones are Grassmann odd, we require 
these Grassmann parity assignations to be compatible with the
multiplication rule $\psi_l\,\psi_{l'}\sim \psi_{l+l'}$. This
compatibility condition induces a $\ZZ_2$ grading on the
charges of the currents. The simplest way to incorporate this
grading is by considering charges that are, in general,
half-integers, \ie\ we shall deal with currents $\psi_l$ with
$2l\in\ZZ$. The fields $\psi_l$ with $l$ integer 
(half-integer) will be considered Grassmann even(odd). If 
$2l\in\ZZ$ we define the function $\epsilon (l)$ as:
\beq
\epsilon (l)\,=\,2(l-[\,l\,])\,,
\label{uno}
\eeq
where $[\,l\,]$ represents the integer part of  $l$, \ie :
\beq
[\,l\,]\,=\,
\cases{l&if $l \in \ZZ$\cr\cr
       l-{1\over 2}&if $l \in \ZZ + {1\over 2}$.}
\label{dos}
\eeq
Notice that $\epsilon (l)$ is zero (one) 
if $l$ is integer (half-integer) and, therefore,  
$\epsilon (l)$ represents the Grassmann parity of the current 
$\psi_l$.

The graded nature of the $\psi_l$ currents is reflected in the
generalized statistics that they obey. In fact, let 
$R(\,\psi_l(z)\psi_{l'}(w)\,)$ denote the radial-ordered
product of the two currents $\psi_l(z)$ and $\psi_{l'}(w)$.
When $|z|\,>\,|w|$ the value of $R(\,\psi_l(z)\psi_{l'}(w)\,)$ is
equal to $\psi_l(z)\psi_{l'}(w)$. In general,  we
shall require that:
\beq
R(\,\psi_l(z)\psi_{l'}(w)\,)\,=\,
(-1)^{\epsilon (l)\,\epsilon (l')}\,\,
{\rm exp}\,\Bigl[\,{2ll'\pi\over k}\,i\,\Bigr]\,
R(\,\psi_{l'}(w)\psi_{l}(z)\,)\,,
\label{tres}
\eeq
where $k$ is a positive integer. Notice that when both $l$ and
$l'$ are integers, eq. (\ref{tres}) is the relation satisfied
by the $Z_k$ parafermionic currents of Zamolodchikov and
Fateev \cite{ZF}. Eq. (\ref{tres}) can also be written as:
\beq
(z-w)^{{2l\,l'\over k}}\,
R(\,\psi_l(z)\psi_{l'}(w)\,)\,=\,
(-1)^{\epsilon (l)\,\epsilon (l')}\,\,
(w-z)^{{2l\,l'\over k}}\,
R(\,\psi_{l'}(w)\psi_{l}(z)\,)\,.
\label{cuatro}
\eeq
It is interesting to point out  that, in order to obtain eq.
(\ref{tres}) from eq. (\ref{cuatro}), we have to put in the
latter $w-z\,=\,e^{i\pi}\,(z-w)$.

The currents $\psi_l$ are primary fields with respect to the
energy-momentum tensor $T$ of the theory. This condition
implies the following Operator Product Expansion (OPE)
\footnote{Although we will not indicate it explicitly, all the 
products of fields appearing in an OPE are radially ordered.}:
\beq
T(z)\,\psi_l(w)\,=\,{\Delta_l\over (z-w)^2}\,\psi_l(w)\,+\,
{\partial \psi_l(w)\over z-w}\,\,,
\label{cinco}
\eeq
where $\Delta_l$ is the conformal weight of the current
$\psi_l$. The values of these weights determine the leading
singularity in the OPE of two currents. Indeed, we must have:
\beq
\psi_l(z)\psi_{l'}(w)\,=\,(\,z\,-\,w\,)^{\Delta_{l+l'}\,-\,
\Delta_{l}\,-\,\Delta_{l'}}\,
\Bigl[\,C_{l,l'}\,\psi_{l+l'}(w)\,+\,\cdots\,\Bigr]\,,
\label{seis}
\eeq
where $C_{l,l'}$ are  constants to be determined. The exchange
relation (\ref{tres}) implies the following equation for these
constants:
\beq
{C_{l,l'}\over C_{l',l}}\,=\,
(-1)^{\epsilon (l)\,\epsilon (l')}\,\,
{\rm exp}\,\Bigl[\,{2ll'\pi\over k}\,i\,\Bigr]\,\,
{\rm exp}\,\Bigl[\, i\pi\,(
\Delta_{l+l'}\,-\,\Delta_{l}\,-\,\Delta_{l'}\,)
\,\Bigr]\,,
\label{siete}
\eeq
which fixes the symmetry properties of the $C_{l,l'}$'s.
If, in particular, we take $l=l'$, the left-hand side of eq. 
(\ref{siete}) is equal to one and, as a consequence, we get:
\beq
2\Delta_{l}\,-\,\Delta_{2l}\,=\,
{2\,l^2\over k}\,+\,\epsilon (l)\,.
\label{ocho}
\eeq
We shall regard eq. (\ref{ocho}) as a relation whose
fulfillment determines the values of the weights $\Delta_l$.
Notice that the charge $l$ can take positive and negative
values. We are going to require that 
$\psi_l$ and $\psi_{-l}$ have the same conformal weight, namely:
\beq
\Delta_{-l}\,=\,\Delta_l\,.
\label{nueve}
\eeq
Moreover, the field $\psi_0$ should be identified with the unit
operator and, therefore, we must have $\Delta_0\,=\,0$. With
these conditions it is not difficult to find a solution for eq. 
(\ref{ocho}):
\beq
\Delta_l\,=\,[\,|l|+{1\over 2}\,]\,-\,{l^2\over k}\,.
\label{diez}
\eeq
Again, for integer charges, we recover the set of conformal
weights found in ref. \cite{ZF}. As in \cite{ZF}, the conformal
dimension vanishes for $l=k$, \ie:
\beq
\Delta_k\,=\,0\,.
\label{once}
\eeq
Therefore, we should identify the currents $\psi_{\pm k}$ with
the unit operator. Based on this observation, we can 
restrict the charge $l$ to the range 
$l\,=-k+{1\over 2},\,\cdots\,,0,\cdots, k-{1\over 2}\,\,$ 
$(2l\in\ZZ$) and, thus, we are going to have $4k-1$ currents (
$2k-1$ ``bosonic" and $2k$ ``fermionic"). It is also interesting
to point out that the $\Delta_l$'s satisfy the following
periodicity relation:
\beq
\Delta_{k-l}\,=\,\Delta_l\,,
\,\,\,\,\,\,\,\,\,\,\,\,\,\,\,\,\,\,\,
{\rm for}\,\,\,0\,\le\,l\,\le\,k\,.
\label{doce}
\eeq
Once the values of the conformal weights $\Delta_l$ are known,
the OPEs (\ref{seis}) among the different currents can be written
as:

\bear
\psi_l(z)\,\psi_{l'}(w)\,&=&\,
{C_{l,l'}\over (z-w)^{{2\,l\,l'\over k}\,+\,\epsilon
(l)\,\epsilon (l')}}\,
\Bigl[\, \psi_{l+l'}\,(w)\,+\,\cdots\,\Bigr]\,,
\,\,\,\,\,\,\,\,\,\,\,\,
l+l'\,<\,k\,,
\rc\rc\rc
\psi_l(z)\,\psi_{-l'}(w)\,&=&\,
{C_{l,-l'}\over (z-w)^{{2\,l'\,(k-l)\over k}\,+\,\epsilon
(l)\,\epsilon (l')}}\,
\Bigl[\, \psi_{l-l'}\,(w)\,+\,\cdots\,\Bigr]\,,
\,\,\,\,\,\,\,\,\,\,\,\,
l'\,<\,l\,,
\label{trece}\\ \rc\rc
\psi_l(z)\,\psi_{-l'}(w)\,&=&\,
{C_{l,-l'}\over (z-w)^{{2\,l\,(k-l')\over k}\,+\,\epsilon
(l)\,\epsilon (l')}}\,
\Bigl[\, \psi_{l-l'}\,(w)\,+\,\cdots\,\Bigr]\,,
\,\,\,\,\,\,\,\,\,\,\,\,
l'\,>\,l\,.
\rc\nonumber
\eear

In order to determine completely the parafermionic current algebra
it would remain to  obtain the value of the structure constants 
$C_{l,l'}$. This task will be performed in section 5 by means of
a free field realization of the algebra. In the present section
we shall limit ourselves to derive some general properties of
these constants. First of all, in order to fix uniquely the
values of the $C_{l,l'}$'s, we must adopt a normalization
condition for the currents. We shall normalize the $\psi_l$
operators in such a way that:
\beq
\psi_l(z)\,\psi_{-l}(w)\,=\,
{1\over (z-w)^{{2\,l\,(k-l)\over k}\,+\,\epsilon(l)}}\,
+\,\cdots\,,
\,\,\,\,\,\,\,\,\,\,\,\,\,\,\,\,
\,\,\,\,\,\,\,\,\,\,\,\,\,\,\,\,
l\,\ge\,0\,.
\label{catorce}
\eeq

Moreover, by using the values (\ref{diez}) in eq. (\ref{siete}),
one can easily obtain the following symmetry properties of the
structure constants:

\beq
C_{l,l'}\,=\,(\,-1\,)^{|\,l+\,l'\,|\,-\,|\,l\,|\,-\,|\,l'\,|}\,\,
C_{l',l}\,.\label{quince}
\eeq
As we mentioned above, in order to be more explicit we must find
an explicit realization of the algebra under consideration. This
realization can be found by relating our system with some coset
theory constructed from a CFT based on a Kac-Moody
(super)algebra. We are now going to argue that our graded
parafermions are realized as the coset 
${\rm osp}(1\vert 2)/ {\rm U} (1)$. The osp$(1\vert 2)$ current
algebra is generated by three bosonic currents  $J_{\pm}$ and
$H$ (which close an su$(2)$ algebra) together with two fermionic
operators $j_{\pm}$. The $H$ current corresponds to the Cartan
subalgebra. This last operator can be realized in terms of the
derivative of a scalar field $\varphi$.   The dependence  of
$J_{\pm}$ and $j_{\pm}$ on the Cartan field $\varphi$ can be
derived from their $H$-charges ($\pm 1$ and $\pm {1\over 2}$, 
respectively). After extracting this $\varphi$ dependence from 
$J_{\pm}$ and $j_{\pm}$ we get some operators, which we shall
identify with $\psi_{\pm 1}$ and $\psi_{\pm{1\over 2}}$, 
respectively:
\bear
J_{\pm}\,&=&\,\sqrt{k}\,\psi_{\pm 1}\,e^{\pm i 
\sqrt{{2\over k}}\,
\varphi}\,,\rc
H\,&=&\,i\,\sqrt{{k\over 2}}\partial \varphi\,,\label{dseis}\\
j_{\pm}\,&=&\,\sqrt{2k}\,\psi_{\pm{1\over 2}}\,
e^{\pm {i \over \sqrt{2k}}\,\varphi}\,.\rc
\nonumber
\eear
The factors $\sqrt{k}$ and $\sqrt{2k}$ appearing in the first and
third equations in (\ref{dseis}) are included to fulfill the
normalization condition (\ref{catorce}) (we are adopting the
conventions of ref. \cite{osp} for the osp$(1\vert 2)$ currents).
Notice  that now the positive integer $k$ is identified with the
level of the osp$(1\vert 2)$ current algebra. The energy-momentum
tensor
$T^J$  of the  osp$(1\vert 2)$ theory can be obtained by means
of the Sugawara construction. The corresponding expression is
given by:

\bear
T^J(z)\,=\,{1\over 2k+3}\,&:\,[\,2\,(H(z))^2\,+\,
J^+(z)\,J^-(z)\,+\,
J^-(z)\,J^+(z)\,\rc
&-\,{1\over 2}\,j^+(z)\,j^-(z)\,+\,
{1\over 2}\,j^-(z)\,j^+(z)\,]:\,\,,
\label{dsiete}
\eear
where the double dots denote normal-ordering. By construction, the
currents $J_{\pm}$, $H$ and $j_{\pm}$ are primary fields with
dimension one with respect to $T^J$. The parafermionic
energy-momentum tensor $T$ can be obtained by subtracting the
contribution of the field $\varphi$ from $T^J$. By substituting
eq. (\ref{dseis}) in eq. (\ref{dsiete}), one can verify that:
\beq
T^J\,=\,T\,-\,{1\over 2}\,(\,\partial\varphi\,)^2\,,
\label{docho}
\eeq
\ie\ the $\varphi$ field contributes to $T^J$ as a free scalar
field without background charge. From eq. (\ref{docho}) one can
readily obtain the dimensions of the exponentials of $\varphi$
appearing in eq. (\ref{dseis}) and, thus, the dimensions of 
$\psi_{\pm 1}$ and $\psi_{\pm{1\over 2}}$. The result is:
\beq
\Delta(\,\psi_{\pm 1}\,)\,=\,1\,-\,{1\over k}\,\,,
\,\,\,\,\,\,\,\,\,\,\,\,\,\,\,\,\,\,\,\,\,\,
\Delta(\,\psi_{\pm {1\over 2}}\,)\,=\,
1\,-\,{1\over 4k}\,\,,
\label{dnueve}
\eeq
which, indeed, coincide with the values given in eq.
(\ref{diez}). This result confirms our identification of the
parafermionic theory and the ${\rm osp}(1\vert 2)/ {\rm U} (1)$
coset. Moreover, as the conformal anomaly of the 
${\rm osp}(1\vert 2)$ current algebra is $2k/(2k+3)$, 
we can  get the central charge of the parafermionic theory
by subtracting the contribution of the Cartan boson $\varphi$:
\beq
c\,=\,-{3\over 2k+3}\,.
\label{veinte}
\eeq
Notice that, as $k$ is a positive integer,  $c$ is always
negative. Actually, the ${\rm osp}(1\vert 2)$ theory is not
unitary \cite{osp} and, thus, we do not expect  the 
${\rm osp}(1\vert 2)/ {\rm U} (1)$ coset to be free of negative
norm states. However, as we shall check in next section, the
Hilbert space of our graded parafermionic system has a
representation theory with good truncation properties which is, in
many senses, similar to the one of the ordinary $Z_k$
parafermions. 

From the OPEs of the ${\rm osp}(1\vert 2)$ currents  and
the relation (\ref{dseis}), we can obtain the first values of the
structure constants of the algebra. These values are:
\bear
C_{{1\over 2}, {1\over 2}}\,=&\,{1\over \sqrt k}\,,
\,\,\,\,\,\,\,\,\,\,\,\,\,\,\,\,\,\,\,
C_{-{1\over 2}, -{1\over 2}}\,=\,-{1\over \sqrt k}\,,
\rc\rc
C_{1, -{1\over 2}}\,=&\,-{1\over \sqrt k}\,,
\,\,\,\,\,\,\,\,\,\,\,\,\,\,\,\,\,\,\,
C_{-1, {1\over 2}}\,=\,-{1\over \sqrt k}\,.
\label{vuno}\\
\nonumber
\eear
In principle, by using the associativity condition of the
algebra, one could get the values of the structure constants from
the values given in eq. (\ref{vuno}). However, we shall not
follow this approach here. Instead, we shall postpone the
determination of the constants $C_{l,l'}$ until section 5, where,
in addition to the associativity condition, we shall make use of
a free field representation of the algebra. 

To finish  this section, let us obtain a relation which will be
very useful in section 3. Let us write the first two terms of the
OPEs $\psi_{{1\over 2}}\,(z)\,\psi_{-{1\over 2}}\,(w)$ and 
$\psi_{1}\,(z)\,\psi_{-1}\,(w)$ as:
\bear
\psi_{{1\over 2}}\,(z)\,
\psi_{-{1\over 2}}\,(w)\,&=&\,
(z-w)^{{1\over 2k}-2}\,+\,(z-w)^{{1\over 2k}}\,\,
{\cal O}^{({1\over 2})}\,(w)\,+\,\cdots\,,\rc\rc
\psi_{1}\,(z)\,\psi_{-1}\,(w)\,&=&\,
(z-w)^{{2\over k}-2}\,+\,(z-w)^{{2\over k}}\,\,
{\cal O}^{(1)}\,(w)\,+\,\cdots\,.
\label{vdos}\\
\nonumber
\eear
In eq. (\ref{vdos}),  ${\cal O}^{({1\over 2})}$ and  
${\cal O}^{(1)}$ are  dimension-two operators whose explicit
expression we do not know. However, these operators contribute to
the finite part of  the product of currents appearing in the
Sugawara expression of $T^J$. Indeed, making use of eqs. 
(\ref{dseis}) and (\ref{vdos}) one can evaluate $T^J$ and, after
comparing the result with eq. (\ref{docho}), one concludes that:

\beq
{\cal O}^{(1)}\,-\,{\cal O}^{({1\over 2})}\,=\,
{2k+3\over 2k}\,\,T\,,
\label{vtres}
\eeq
which is the relation we wanted to obtain.

\medskip
\setcounter{equation}{0}
\section{Parafermionic Hilbert Space}
\medskip

In this section we are going to analyze the structure of the
Hilbert space of the parafermionic theory introduced in the
previous section. First of all, it is clear that the charge
structure of the model induces a decomposition of its Hilbert
space $\cal H$. Let us denote by $p$ and $\bar p$ the left and
right charges respectively and let ${\cal H}_{(p,\bar p)}$ be the
subspace of $\cal H$ with the indicated values of the charges
\footnote{We shall adopt the units of ref. \cite{ZF} for the left
and right charges.}.
The Hilbert space $\cal H$ splits into a direct sum of the type:

\beq
{\cal H}\,=\,\oplus\,{\cal H}_{(p,\bar p)}\,.
\label{vcuatro}
\eeq
As  is well-known, in CFT there is a one-to-one correspondence
between states in the Hilbert space and operators. Therefore, we
shall consider $\cal H$ also as the field space of the theory and
eq. (\ref{vcuatro}) will be regarded as the decomposition of the
space of fields according to the different charge sectors of the
theory. 

Notice that the left and right charges $p$ and $\bar p$ of the
parafermionic current $\psi_l$ are $p=\,2l$ and $\bar
p\,=\,0$ (\ie\ $\psi_l\,\in\,{\cal H}_{(2l,0)}$, 
see ref. \cite{ZF} for details). The so-called mutual locality
exponent $\gamma$ of two fields is defined  as the phase, in
units of $2\pi$, that is generated when we circle one field
around the other inside a correlation function. The charges of
the fields determine the value of $\gamma$ \cite{ZF}. Indeed, let 
$\phi_{(p_1,\bar p_1)}$ and $\phi_{(p_2,\bar p_2)}$ be two
arbitrary fields 
($\phi_{(p_i,\bar p_i)}\,\in\, {\cal H}_{(p_i,\bar p_i)}$). Their
mutual locality exponent is given by:
\beq
\gamma\,\Bigl(\,\phi_{(p_1,\bar p_1)}\,,\,
\phi_{(p_2,\bar p_2)}\,\Bigr)\,=\,
-{1\over 2k}\,
[\,p_1\,p_2\,-\,\bar p_1\,\bar p_2\,]\,,
\,\,\,\,\,\,\,\,\,\,\,\,\,\,\,\,
{\rm mod}\,\,\ZZ\,.
\label{vcinco}
\eeq
Notice that $\gamma$ is defined modulo $\ZZ$ and it does not
change if any of the $p_i$ or $\bar p_i$ is shifted by $2k$. It is
also important to point out that 
$\gamma\,\Bigl(\,\phi_{(p_1,\bar p_1)}\,,\,
\phi_{(p_2,\bar p_2)}\,\Bigr)$ determines the non-local part of
the OPE $\phi_{(p_1,\bar p_1)}(z)\,\,
\phi_{(p_2,\bar p_2)}(w)$. Let us check this fact in the case of
the parafermionic currents. Indeed,  as 
$\psi_l\,\in\,{\cal H}_{(2l,0)}$), we have:
\beq
\gamma\,\Bigl(\,\psi_l\,,\,
\psi_{l'}\,\Bigr)\,=\,-{2ll'\over k}\,,
\label{vseis}
\eeq
in agreement with the parafermionic OPEs (\ref{trece}). Notice
that if $\phi_{(p,\bar p)}\,\in\, {\cal H}_{(p,\bar p)}$,  then
$\psi_l\,\phi_{(p,\bar p)}\,\in\, {\cal H}_{(p+2l,\bar p)}$.
Moreover, from eq. (\ref{vcinco}) one has:
\beq
\gamma\,\Bigl(\,\psi_l\,,\,
\phi_{(p,\bar p)}\,\Bigr)\,=\,-{lp\over k}\,,
\label{vsiete}
\eeq
and, therefore, it is clear that the currents $\psi_1$ and 
$\psi_{-1}$  act on the fields $\phi_{(l,\bar l)}$ according to
the general operator expansion:
\beq
\psi_{\pm 1}\,(z)\,\phi_{(l,\bar l)}\,(0)\,=\,
\sum_{m\,=\,-\infty}^{+\infty}\,\,
z^{\mp{l\over k}\,+\,m-1}\,A_{\pm {l\over k}-m}^{(\pm)}\,
\phi_{(l,\bar l)}\,(0)\,.
\label{vocho}
\eeq
It is clear from our previous discussion that :
\beq
A_{\pm{l\over k}-m}^{(\pm)}\,\phi_{(l,\bar l)}\,\in\,
{\cal H}_{l\pm2,\bar l}\,.
\label{vnueve}
\eeq
Furthermore, if $h$ is the (holomorphic) conformal weight of the
field $\phi_{(l,\bar l)}$,  one easily
concludes from eq. (\ref{dnueve}) that the
dimensions of the fields appearing on  the right-hand side of
eq.~(\ref{vocho}) are:
\beq
\Delta\,\Bigl(\,A_{\pm{l\over k}-m}^{(\pm)}
\,\phi_{(l,\bar l)}\,\Bigr)
\,=\,h\,+\,m\,-\,{1\pm l\over k}\,\,.
\label{treinta}
\eeq
Similarly, the action of the currents $\psi_{{1\over 2}}$
and $\psi_{-{1\over 2}}$ on the fields $\phi_{(l,\bar l)}$ is given
by:
\beq
\psi_{\pm{1\over 2}}\,(z)\,\phi_{(l,\bar l)}\,(0)\,=\,
\sum_{m\,=\,-\infty}^{+\infty}\,\,
z^{\mp{l\over 2k}\,+\,m-1}
\,B_{\pm{l\over 2k}-m}^{(\pm)}\,
\phi_{(l,\bar l)}\,(0)\,,
\label{tuno}
\eeq
where now:
\beq
B_{\pm{l\over 2k}-m}^{(\pm)}
\,\phi_{(l,\bar l)}\,\in\,
{\cal H}_{l\pm 1,\bar l}\,,
\,\,\,\,\,\,\,\,\,\,\,\,\,\,\,\,\,\,\,
\Delta\,\Bigl(\,B^{(\pm)}_{\pm {l\over 2k}-m}\,
\,\phi_{(l,\bar l)}\, \Bigr)
\,=\,h\,+\,m\,-\,{1\pm 2l\over 4k}\,.
\label{tdos}
\eeq
The action of the mode operators on the different fields can be
represented as a  contour integral. Indeed, it follows from eqs. 
(\ref{vocho}) and (\ref{tuno}) that one can write:
\bear
A^{(\pm)}_{\pm{l\over k}+m}\,\phi_{(l,\bar l)}(0)\,&=&\,
\oint_C\,{dz\over 2\pi i}\,z^{\pm{l\over k}+m}\,
\psi_{\pm 1}\,(z)\, \phi_{(l,\bar l)}(0)\,\,,
\rc\rc
B^{(\pm)}_{\pm{l\over 2k}+m}\,\phi_{(l,\bar l)}(0)\,&=&\,
\oint_C\,{dz\over 2\pi i}\,z^{\pm{l\over 2k}+m}\,
\psi_{\pm{1\over 2}}\,(z)\, \phi_{(l,\bar l)}(0)\,\,.
\label{ttres}\\
\nonumber
\eear
This representation will be very useful in what follows.

The mode operators of eqs. (\ref{vocho}) and (\ref{tuno}) satisfy
a series of generalized (anti)commutation relations which we are
now going to derive. These relations define what is called  a $Z$
algebra in the
mathematical literature. Exploiting this algebra we will be able
to uncover the general structure of the representation theory of
the graded parafermionic model. In order to obtain these $Z$
algebra relations we follow closely the method of ref. \cite{ZF}.
Let us consider the integrals: 
\beq
\oint_{C_1}\,{dz_1\over 2\pi i}\,
\oint_{C_2}\,{dz_2\over 2\pi i}\,\,
R\,\Bigl(\,
\psi_{{1\over 2}}\,(z_1)\,
\psi_{-{1\over 2}}\,(z_2)\,
\Bigr)\,
z_1^{{l\over 2k}+n}\,\,z_2^{-{l\over 2k}+m+1}\,
(\,z_1\,-\,z_2\,)^{-{2k+1\over 2k}}\,\,
\phi_{(l,\bar l)}(0)\,.
\label{tcuatro}
\eeq
where $n$ and $m$ are integers. The powers of $z_1$ and $z_2$
have been chosen to make the integrand in (\ref{tcuatro}) single
valued. We first evaluate the integrals in (\ref{tcuatro}) along
contours $C_1$ and $C_2$ such that $|\,z_1\,|\,>\,|\,z_2\,|$,
\ie\ such that $C_2$ lies inside $C_1$. In this case, the radial
ordering does not modify the order in which the fields are written
in eq. (\ref{tcuatro}). The term 
$(\,z_1\,-\,z_2\,)^{-{2k+1\over 2k}}$ can be expanded  in powers
of $z_2/z_1$  by using the equation:
\beq
(\,1\,-\,x\,)^{\lambda}\,=\,\sum_{r=0}^{+\infty}\,
{\cal D}_{\lambda}^{(r)}\,x^r\,,
\label{tcinco}
\eeq
where the coefficients ${\cal D}_{\lambda}^{(r)}$ are given by:
\beq
{\cal D}_{\lambda}^{(r)}\,=\,
{\Gamma\,(\,r\,-\,\lambda\,)\over r!\,
\Gamma\,(\,-\lambda\,)}\,=\,
{(-1)^r\over  r!}\,\,
\prod_{j=0}^{r-1}\,(\,\lambda\,-\,j\,)\,.
\label{tseis}
\eeq
\begin{figure}
\centerline{\hskip -.8in \epsffile{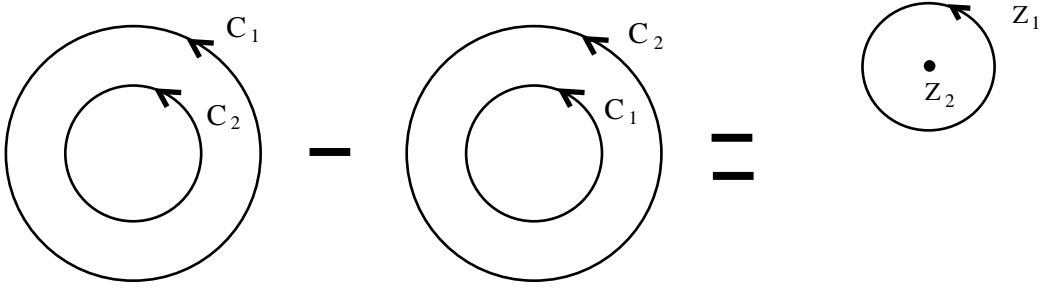}}
\caption{Diagrammatic representation of 
the contour manipulation needed to obtain the generalized
(anti)commutation relations of  the mode operators.}
\label{fig1}
\end{figure}
Making use of eqs. (\ref{tcinco}) and (\ref{ttres}), the integral 
(\ref{tcuatro}) can be computed for the class of contours
considered. Obviously, we could apply the same procedure when the
contour $C_1$ lies inside $C_2$. Notice that in this case the
radial ordering reverses the order in which the fields are
multiplied in (\ref{tcuatro}). Actually, the
radially ordered product can be computed by means of the relation 
(\ref{cuatro}). As shown in figure 1, the difference between the
integrals (\ref{tcuatro}),  evaluated for the two types of contours
described above,  gives rise to an integral in which $z_1$ is
integrated along a small contour centered around $z_2$. The value
of this last integral can be obtained by making use of the
parafermionic OPEs (\ref{vdos}). The final result of the
calculation is the following algebraic relation satisfied by the
modes of $\psi_{{1\over 2}}$ and $\psi_{-{1\over 2}}$:
\bear
\sum_{r=0}^{+\infty}\,{\cal D}_{-{2k+1\over 2k}}^{(r)}\,
&\Bigl[\,B^{(+)}_{{l-1\over 2k}\,+\,n-r-1}\,\,
B^{(-)}_{-{l\over 2k}\,+\,m+r+1}\,\,-\,\,
B^{(-)}_{-{l+1\over 2k}\,+\,m-r}\,\,
B^{(+)}_{{l\over 2k}\,+\,n+r}\,\,\Bigr]\,=\,\rc
&=\,{1\over 2}\,(\,{l\over 2k}\,+\,n\,)\,
\,(\,{l\over 2k}\,+\,n\,-\,1\,)\,
\delta_{n+m,0}\,+\,
{\cal O}^{({1\over 2})}_{n+m}\,\,.\label{tsiete}\\
\nonumber
\eear
Similarly,  we could get the relation verified  by the modes of
$\psi_{1}$ and $\psi_{-1}$. The result is:
\bear
\sum_{r=0}^{+\infty}\,{\cal D}_{-{k+2\over k}}^{(r)}\,
&\Bigl[\,A^{(+)}_{{l-2\over k}\,+\,n-r-1}\,\,
A^{(-)}_{-{l\over k}\,+\,m+r+1}\,\,+\,\,
A^{(-)}_{-{l+2\over  k}\,+\,m-r}\,\,
A^{(+)}_{{l\over k}\,+\,n+r}\,\,\Bigr]\,=\,\rc
&=\,{1\over 2}\,(\,{l\over k}\,+\,n\,)\,
\,(\,{l\over k}\,+\,n\,-\,1\,)\,
\delta_{n+m,0}\,+\,
{\cal O}^{(1)}_{n+m}\,\,.\label{tocho}\\
\nonumber
\eear
In eqs. (\ref{tsiete}) and (\ref{tocho}),  
${\cal O}^{({1\over 2})}_{n}$ and ${\cal O}^{(1)}_{n}$ are the
modes of the operators appearing on the right-hand side of eq.
(\ref{vdos}), namely:
\beq
{\cal O}^{(\alpha)}_{n}\,=\,\oint\,\,
{dz\over 2\pi i}\,\,z^{n+1}\,{\cal O}^{(\alpha)}(z)\,\,,
\,\,\,\,\,\,\,\,\,\,\,\,\,\,\,\,\,\,\,\,\,
\,\,\,\,\,\,\,\,\,\,\,\,\,\,\,\,\,\,\,\,\,
\alpha\,=\, {1\over 2}\,,\,1\,.
\label{tnueve}
\eeq
Before extracting some consequences from eqs. (\ref{tsiete}) and
(\ref{tocho}), let us write down some other $Z$-algebraic
relations obtained by applying the technique just described.
First of all, from the integrals:

\beq
\oint_{C_1}\,{dz_1\over 2\pi i}\,
\oint_{C_2}\,{dz_2\over 2\pi i}\,\,
R\,\Bigl(\,
\psi_{\pm{1\over 2}}\,(z_1)\,
\psi_{\pm{1\over 2}}\,(z_2)\,
\Bigr)\,
z_1^{\pm{l\over 2k}+n}\,\,z_2^{\pm{l\over 2k}+m}\,
(\,z_1\,-\,z_2\,)^{{1\over 2k}}\,\,
\phi_{(l,\bar l)}(0)\,,
\label{cuarenta}
\eeq
we get:

\beq
\sum_{r=0}^{+\infty}\,{\cal D}_{{1\over 2k}}^{(r)}\,
\Bigl[\,B^{(\pm)}_{{1\pm l\over 2k}\,+\,n-r}\,\,
B^{(\pm)}_{\pm{l\over 2k}\,+\,m+r}\,\,+\,\,
B^{(\pm)}_{{1\pm l\over 2k}\,+\,m-r}\,\,
B^{(\pm)}_{\pm{l\over 2k}\,+\,n+r}\,\,\Bigr]\,
=\,\pm\,{1\over \sqrt k}\,A^{(\pm)}_{\pm{l\over k}\,+\,n+m}\,.
\label{cuno}
\eeq
Similarly, if we consider the integrals:

\beq
\oint_{C_1}\,{dz_1\over 2\pi i}\,
\oint_{C_2}\,{dz_2\over 2\pi i}\,\,
R\,\Bigl(\,
\psi_{\pm1}\,(z_1)\,
\psi_{\mp{1\over 2}}\,(z_2)\,
\Bigr)\,
z_1^{\pm{l\over k}+n}\,\,z_2^{\mp{l\over 2k}+m}\,
(\,z_1\,-\,z_2\,)^{-{1\over k}}\,\,
\phi_{(l,\bar l)}(0)\,,
\label{cdos}
\eeq
the following relation is obtained:

\beq
\sum_{r=0}^{+\infty}\,{\cal D}_{-{1\over k}}^{(r)}\,
\Bigl[\,A^{(\pm)}_{-{1\mp l\over k}\,+\,n-r}\,\,
B^{(\mp)}_{\mp{l\over 2k}\,+\,m+r}\,\,-\,\,
B^{(\mp)}_{-{2\pm l\over 2k}\,+\,m-r}\,\,
A^{(\pm)}_{\pm{l\over k}\,+\,n+r}\,\,\Bigr]\,
=\,-{1\over \sqrt k}\,
B^{(\pm)}_{\pm{l\over 2k}\,+\,n+m}\,.
\label{ctres}
\eeq

Notice that, in order to evaluate the right-hand side of eqs. 
(\ref{cuno}) and (\ref{ctres}), the values of the structure
constants (\ref{vuno}) are needed.

The primary fields of the parafermionic algebra are those that
satisfy a certain highest-weight condition. Let us denote them by
$\Phi^{l, \bar l}_{l, \bar l}$, where $l$ and $\bar l$ refer
respectively to the holomorphic and antiholomorphic quantum
numbers. All the fields in the parafermionic module are obtained as
descendants of the $\Phi^{l, \bar l}_{l, \bar l}$'s. 
The highest-weight conditions satisfied by 
$\Phi^{l, \bar l}_{l, \bar l}$ are:
\bear
&B^{(+)}_{{l\over 2k}\,+\,n}\,\,
\Phi^{l, \,\bar l}_{l, \,\bar l}\,=\,
A^{(+)}_{{l\over k}\,+\,n}\,\,
\Phi^{l, \,\bar l}_{l, \,\bar l}\,=\,0\,,
\,\,\,\,\,\,\,\,\,\,\,\,\,\,\,\,\,\,\,\,\,\,\,\,\,\,\,
\,\,\,\,\,\,\,\,\,\,\,\,\,
n\ge 0\,,\rc\rc
&B^{(-)}_{-{l\over 2k}\,+\,n\,+1}
\,\,\Phi^{l, \,\bar l}_{l, \,\bar l}\,=\, 
A^{(-)}_{-{l\over k}\,+\,n\,+1}\,\,
\Phi^{l,\, \bar l}_{l, \,\bar l}\,=\,0\,,
\,\,\,\,\,\,\,\,\,\,\,\,\,\,\,\,\,\,\,\,n\ge 0\,,
\label{ccuatro}\\
\nonumber
\eear
together with analogous antiholomorphic relations. Notice that
the requirement (\ref{ccuatro})  implies that the dimensions of
the fields belonging to the module constructed by acting on 
$\Phi^{l, \bar l}_{l, \bar l}$ with ``creation" operators are
bounded from below.

The algebraic relations (\ref{tsiete}) and (\ref{tocho}) for
$n=m=0$ determine the dimensions of the primary fields. Indeed,
applying in this case eqs. (\ref{tsiete}) and (\ref{tocho}) to 
$\Phi^{l, \bar l}_{l, \bar l}$, we see that the left-hand side of
the resulting equations vanish as a consequence of the
highest-weight conditions (\ref{ccuatro}). Moreover, by subtracting
these two equations we generate on the right-hand side the
energy-momentum tensor due to eq. (\ref{vtres}). Taking into
account that the zero mode of the energy-momentum tensor acts
diagonally on the field $\Phi^{l, \bar l}_{l, \bar l}$ and its
eigenvalue is precisely the conformal dimension $h_l$ of   
$\Phi^{l, \bar l}_{l, \bar l}$, one gets:
\beq
h_l\,=\,{l(2k-3l)\over 4k(2k+3)}\,\,.
\label{ccinco}
\eeq

It will be argued in section 4 that, from the analysis of the
three-point functions of the model, the values of the
highest-weight charges are restricted to the range $l\le k$. From
now on in this section, we shall assume that $l$ satisfies this
condition. It is important to notice that the dimensions $h_l$ in
eq.  (\ref{ccinco}) can become negative within this range of
values of $l$ (recall that the theory we 
are dealing with is not unitary).

In the remaining of this section we shall suppress, for
notational simplicity, the dependence of the fields on the
antiholomorphic quantum numbers and so, for example, we shall
write $\Phi_l^l$ instead of $\Phi^{l, \bar l}_{l, \bar l}$. All
the fields in the parafermionic module can be obtained by applying
the mode operators to $\Phi_l^l$. Let us determine the conformal
dimensions of the fields generated in this way. First of all, we
consider the fields
\beq
\Phi^{l}_{l+n}\,=\,
B^{(+)}_{{l+n-1\over 2k}\,-1}\,
\cdots\,B^{(+)}_{{l+1\over 2k}\,-1}\, 
B^{(+)}_{{l\over 2k}\,-1}\,\Phi^{l}_{l}\,,
\label{cseis}
\eeq
where $n\ge 0\,\,,\,\,n\in \ZZ$. Notice that
$\Phi^{l}_{l+n}\,\in\,{\cal H}_{(\,l+n,\bar l\,)}$. The dimension
of $\Phi^{l}_{l+n}$  can be computed from
(\ref{tdos}) and the result is:
\beq
h^l_{l+n}\,\equiv\,
\Delta\,(\,\Phi^{l}_{l+n}\,)\,=\,h_l\,+\,
{n\over 4k}\,(\,4k\,-\,2l\,-\,n\,)\,.
\label{csiete}
\eeq
Similarly, we can define the fields $\Phi^{l}_{l-n}$:
\beq
\Phi^{l}_{l-n}\,=\,
B^{(-)}_{-{l-n+1\over 2k}}\,\cdots
\,B^{(-)}_{-{l-1\over 2k}\,}\,
B^{(-)}_{-{l\over 2k}}\,\Phi^{l}_{l}\,,
\label{cocho}
\eeq
where, again, $n$ is a non-negative integer and now 
$\Phi^{l}_{l-n}\,\in\,{\cal H}_{(\,l-n,\bar l\,)}$. 
The calculation of
the conformal weight in this case gives:

\beq
h^l_{l-n}\,\equiv\,
\Delta\,(\,\Phi^{l}_{l-n}\,)\,=\,h_l\,+\,
{n\over 4k}\,(\,2l\,-\,n\,)\,.
\label{cnueve}
\eeq

In a parafermionic highest-weight module the descendant fields
have dimensions which, in general, do not differ by an integer.
Actually, in our case, the fields whose conformal weights have
different fractional part belong to  a finite subset, which can be
selected by restricting appropriately the range of values of the
charge. Indeed, let $h_m^l$ be the dimension of $\Phi^{l}_{m}$.
For $m\ge l$ ($m\le l$),  $h_m^l$ is given by eq. (\ref{csiete}) 
(eq. (\ref{cnueve}) respectively). 
In principle,  $m$ can be any
integer. However, it is easy to  verify from eqs.
(\ref{csiete})  and (\ref{cnueve}) that:
\bear
h_{m+2k}^{l}&=&h_{m}^{l}\,+\,k-m\,,
\,\,\,\,\,\,\,\,\,\,\,\,\,\,\,\,\,\,
m\ge l\,,\rc\rc
h_{m-2k}^{l}&=&h_{m}^{l}\,+\,m-k\,,
\,\,\,\,\,\,\,\,\,\,\,\,\,\,\,\,\,\,
m\le l\,,\label{cincuenta}\\
\nonumber
\eear
\ie\ shifting the charge by $2k$ ($-2k$)  in eq. (\ref{csiete})
(eq. (\ref{cnueve})), the corresponding dimension is shifted by
an integer. Therefore, in the determination of the values of 
$h_m^l$ modulo $\ZZ$,  we can restrict $m$ to the range 
$l-2k\le m \le 2k+l$. Actually, the upper limit of the previous
interval can be refined,  since it is straightforward to check from
eqs.  (\ref{csiete}) and (\ref{cnueve}) that:
\beq
h_{2k+l-r}^{l}\,=\,h^{l}_{l-r}\,+\,k-l\,,
\,\,\,\,\,\,\,\,\,\,\,\,\,\,\,\,\,\,
0\le r\le 2l\,.
\label{ciuno}
\eeq
Eq. (\ref{ciuno}) implies that we can take 
$l-2k\le m \le 2k-l$. Moreover, the values of $h_m^l$ for 
$l-2k\le m \le -l$ differ by an integer from the $h_m^l$'s for
$l\le m \le 2k-l$, namely:
\beq
h_{-l-r}^{l}=h_{2k-l-r}^{l}\,+\,l-k\,,
\,\,\,\,\,\,\,\,\,\,\,\,\,\,\,\,\,\,\,\,\,\,\,\,\,\,\,\,\,
0\le r\le 2k-2l\,.
\label{citres}
\eeq
Taken together,  eqs. (\ref{cincuenta})-(\ref{citres}) imply that
$m$ can be restricted to the range $-l\le m \le 2k-l$ with
$m-l\in\ZZ$. The corresponding dimensions can be written as:
\beq
h^l_m\,=\,\cases{h_l\,+\,{(l-m)(l+m)\over 4k}
                     &if $-l\,\le\,m\le\,l$\cr\cr
                     h_l\,+\,{(m-l)(4k-l-m)\over 4k}
                    &if $l\,<\,m\le\,2k-l$.}
\label{cicuatro}
\eeq

The analysis we have just done shows that all the fields
obtained by acting with the mode operators on $\Phi_l^l$  have
dimensions that differ by integers from the values written in eq. 
(\ref{cicuatro}). Let us now define:
\beq
\hat h^l_m\,=\,{l(l+1)\over 2(2k+3)}\,-\,
{m^2\over 4k}\,.
\label{cicinco}
\eeq
Then   eq. (\ref{cicuatro}) can be put as:
\beq
h^l_m\,=\,\cases{\hat h^l_m
                     &if $-l\,\le\,m\le\,l$\cr\cr
                    \hat h^l_m\,+\,m-l
                    &if $l\,<\,m\le\,2k-l$.}
\label{ciseis}
\eeq
In  next section we shall interpret $\hat h^l_m$ in terms of
the primary fields of the osp$(1\vert 2)$ current algebra. This
interpretation, together with eq. (\ref{ciseis}), will allow us
to give an explicit free field representation of the operators
$\Phi_m^l$, which, making use of the results of ref. \cite{osp},
can be used to compute the correlation functions of the
parafermionic theory.

\medskip
\setcounter{equation}{0}
\section{Free Field Representation}
\medskip

In order to find an explicit free field realization of the graded 
parafermionic symmetry, let us consider again its realization in
the osp$(1\vert 2)/U(1)$ coset CFT. According to eq.
(\ref{dseis}), the generating parafermions $\psi_{\pm 1}$ and 
$\psi_{\pm {1\over 2}}$ are obtained by extracting the Cartan
dependence from the osp$(1\vert 2)$ currents. This factorization
can be neatly done in the framework of a free field realization
of the osp$(1\vert 2)$ CFT, such as the one proposed in ref.
\cite{ber} and further developed in refs. \cite{osp, rama}. Indeed,
after some redefinitions of the fields considered in ref.
\cite{osp}, one can put the osp$(1\vert 2)$ currents in the form
given in eq. (\ref{dseis}). The parafermionic currents are realized
in terms of two scalar fields $\varphi_1$ and $\varphi_2$ and a
pair of conjugate anticommuting ghost fields $\eta$ and $\xi$. The
scalar fields $\varphi_1$ and $\varphi_2$ are normalized in such a
way that their OPE is given by:
\beq
\varphi_i(z)\,\varphi_j(w)\,=\,-\delta_{ij}\,
{\rm log}\,(z-w)\,,
\label{cisiete}
\eeq
whereas the fields  $\eta$ and $\xi$ satisfy:
\beq
\eta (z)\,\xi(w)\,=\,\xi (z)\,\eta(w)\,=\,
{1\over z-w}\,\,,
\label{ciocho}
\eeq
and their dimensions are:
\beq
\Delta(\eta)\,=\,1\,,
\,\,\,\,\,\,\,\,\,\,\,\,\,\,\,\,\,\,\,\,\,\,
\Delta(\xi)\,=\,0\,.
\label{cinueve}
\eeq
In terms of the free fields, the energy-momentum tensor $T$ can be
obtained, through the Sugawara construction, as in eqs.
(\ref{dsiete}) and (\ref{docho}). Its explicit expression is:
\beq
T\,=\,-{1\over 2}\,(\partial\varphi_1)^2\,
-{1\over 2}\,(\partial\varphi_2)^2\,-\,
{i\over 2\sqrt{2k+3}}\,\partial^2\,\varphi_2\,-\,
\eta\partial \xi\,.
\label{sesenta}
\eeq
It is a simple exercise to check that the central charge obtained
from eq. (\ref{sesenta}) coincides with the one written in eq.
(\ref{veinte}). Moreover, in terms of $\varphi_1$, $\varphi_2$, 
$\eta$ and $\xi$, the basic parafermionic currents 
$\psi_{\pm 1}$ and  $\psi_{\pm {1\over 2}}$ are given by:
\bear
\psi_1\,&=&\,\Bigl[\,
{1\over \sqrt 2}\,\partial\varphi_1\,
+\,{i\over 2}\,
\sqrt{{2k+3\over k}}\,\partial\varphi_2\,
+{1\over 2\sqrt{k}}\,\eta\xi\,
\Bigr]\,
e^{\sqrt{{2\over k}}\varphi_1}\,,\rc\rc
\psi_{{1\over 2}}\,&=&\,\Bigl[\,\bigl(\,
{1\over 2}\,\partial\varphi_1\,+\,{i\over 2}\,
\sqrt{{2k+3\over 2k}}\,\partial\varphi_2\,\bigr)\,\xi\,-\,
{1\over 2 \sqrt{2k}}\,\partial \xi\,+\,
{1\over \sqrt{2k}}\,\eta\,\Bigr]\,
e^{{1\over\sqrt{2k}}\varphi_1}\,,\rc\rc
\psi_{-{1\over 2}}\,&=&\,\Bigl[\,\bigl(\,-
{1\over 2}\,\partial\varphi_1\,+\,{i\over 2}\,
\sqrt{{2k+3\over 2k}}\,\partial\varphi_2\,\bigr)\,\xi\,+\,
{4k+3\over 2\sqrt {2k}}\,\partial \xi\,-\,
{1\over \sqrt{2k}}\,\eta\,\Bigr]\,
e^{-{1\over\sqrt{2k}}\varphi_1}\,,\rc\rc
\psi_{-1}\,&=&\,\Bigl[\,
-{1\over \sqrt 2}\,\partial\varphi_1\,
+\,{i\over 2}\,
\sqrt{{2k+3\over k}}\,\partial\varphi_2\,
+{1\over 2\sqrt k}\,\eta\xi\,+{k+1\over \sqrt k}\,\xi\partial\xi\,
\Bigr]\,
e^{-\sqrt{{2\over k}}\varphi_1}\,.\label{suno}\\
\nonumber
\eear

The graded nature of the currents $\psi_{\pm 1}$ and  
$\psi_{\pm {1\over 2}}$ is manifest from their representation in 
eq. (\ref{suno}). After a short calculation it may be verified
that  $\psi_{\pm 1}$ and   $\psi_{\pm {1\over 2}}$ are primary
fields of the Virasoro algebra and that their dimensions are given
by eq. (\ref{dnueve}). Moreover, it is easy to check that they
indeed satisfy the parafermionic algebra of eq. (\ref{trece})
with the structure constants given by eq. (\ref{vuno}). From the
expressions given in eq. (\ref{suno}), one can generate all the
other parafermionic currents of the model and get the structure
constants of the algebra. This will be done in section 5.

The factorization of the Cartan degrees of freedom  in the 
osp$(1\vert 2)$ theory can also  be performed on the primary
fields of the model \cite{osp}. Let $G_m^l$ be an osp$(1\vert 2)$
primary field corresponding to a value $l/2$ of the isospin  and
to a $H$-charge equal to  $m/2$. According to the osp$(1\vert 2)$
representation theory \cite{Pais}, the isospin of the
finite-dimensional representations can take any positive integer
or half-integer value and the difference between the Cartan
eigenvalue and the isospin can be integer or half-integer. This
implies that both $l$ and $m$ in $G_m^l$ are integers.
Moreover \cite{Pais},  in an osp$(1\vert 2)$  multiplet, $m$ can
take values in the range $-l\le m\le l$.  After extracting the
Cartan dependence of $G_m^l$, one gets:
\beq
G_m^l\,=\,\Phi_m^l\,\,exp[\,i {m\over \sqrt{2k}} \,\,
\varphi\,]\,,
\label{sdos}
\eeq
where $\Phi_m^l$ is an operator which only depends on the fields of
the osp$(1\vert 2)/U(1)$ coset. The explicit expression of
$\Phi_m^l$ for $-l\,\le\,m\le l$ is:
\beq
\Phi^l_m\,=\,\cases{exp\,[\,{m\over \sqrt{2 k}}
                    \,\,\varphi_1\,
                     +\,{il\over\sqrt{2k+3}}\,\,
                     \varphi_2\,]  
                     &if $l-m\in 2\,\ZZ$\cr\cr
                     \xi\,exp\,[\,{m\over\sqrt{2 k}}
                     \,\,\varphi_1\,
                     +\,{il\over \sqrt{2k+3}}\,
                    \varphi_2\,]  
                    &if $l-m\in 2\,\ZZ\,+\,1\,$.}
\label{stres}
\eeq

Notice that the Grassmann character of $\Phi_m^l$ depends on
whether $l-m$ is even or odd. From eqs. (\ref{stres}) and 
(\ref{sesenta}), it is immediate to find the conformal dimensions
of the $\Phi_m^l$'s. The result of the calculation is:
\beq
\Delta (\,\Phi_m^l\,)\,=\,{l(l+1)\over 2(2k+3)}\,-\,
{m^2\over 4k}\,=\,\hat h^l_m\,,
\label{scuatro}
\eeq
where $\hat h^l_m$ is the same as in eq. (\ref{cicinco}). In view
of the result displayed in eq. (\ref{ciseis}), one is tempted to
identify the fields (\ref{stres}) with the $\Phi_m^l$'s defined
in eq. (\ref{cocho}) for $-l\,\le\,m\le l$. Indeed, the operators 
(\ref{stres}) for $m=l$ verify the highest-weight conditions  
(\ref{ccuatro}). Moreover, acting with the $\psi_{ -{1\over 2}}$
current on the highest-weight field is equivalent, in the 
osp$(1\vert 2)$ theory, to a multiplication by $j_-$ and, as
shown in ref. \cite{osp}, the different components of the $G_m^l$
operator are generated in this way. Alternatively, one can generate
directly the form of the $\Phi_m^l$'s by using the explicit
expressions  of $\psi_{ \pm {1\over 2}}$ and $\Phi_l^l$,   given
in eqs. (\ref{suno}) and (\ref{stres}) respectively. As the
result of this calculation one can prove that, indeed, for 
$-l\,\le\,m\le l$, the expressions of  $\Phi_m^l$ displayed in eq.
(\ref{stres}) are recovered. For $m>l$ one obtains that $\Phi_m^l$
contains an exponential, such as the one in eq. (\ref{stres}),
multiplied by a polynomial in the derivatives of the fields of
dimension
$m-l$. Notice that this representation of the $\Phi_m^l$'s for 
$m>l$ is in agreement with the dimensions written in eq. 
(\ref{ciseis}) for this case. 

The relation (\ref{sdos}) between operators of the 
osp$(1\vert 2)$ and parafermionic theories allows one to express
the correlation functions of the latter in terms of the vacuum
expectation values of the former. Actually, from eq.
(\ref{sdos}) one obtains the following general result:
\bear
\langle\,\,\Phi_{m_1,\bar m_1}^{l_1,\bar l_1}\,(z_1, \bar
z_1)\,\,\cdots\,\, \Phi_{m_n,\bar m_n}^{l_n,\bar l_n}\,(z_n, \bar
z_n)
\,\,\rangle\,&=&\,
\prod_{i<j}^{n}\,(\,z_i\,-\,z_j\,)^{-{m_i m_j\over 2k}}\,
(\,\bar z_i\,-\,\bar z_j\,)^{-{\bar m_i \bar m_j\over
2k}}\,\times\rc\rc &\times& 
\langle\,\,G_{m_1,\bar m_1}^{l_1,\bar l_1}\,(z_1, \bar
z_1)\,\,\cdots\,\,
G_{m_n,\bar m_n}^{l_n,\bar l_n}\,(z_n, \bar z_n)
\,\,\rangle\,.\rc
\label{scinco}
\eear
As the Cartan degrees
of freedom only contribute in eq. (\ref{scinco}) with  factors
which are powers of the coordinates, it is clear that the
selection rules of the osp$(1\vert 2)$ theory are inherited in
the parafermionic model. These selection rules have been studied
in detail in ref. \cite{osp}. Actually, by studying the operator
product algebra of the osp$(1\vert 2)$ CFT, it was demonstrated in
ref.~\cite{osp} that only those isospins lower or equal than $k/2$
are coupled ($k$ being the level of the osp$(1\vert 2)$ Kac-Moody
algebra). This result implies that, as was advanced around eq.
(\ref{ccinco}), the charges $l$ of the highest-weight
parafermionic fields
$\Phi_l^l$ are restricted by the condition:
\beq
l\,\le\,k\,.
\label{sseis}
\eeq
Moreover, the analysis of ref. \cite{osp} can be used to get some
highly non-trivial results in the parafermionic theory. Let us
see, as an example, what is the value of the vacuum expectation
value of the product of three  highest-weight parafermionic
fields. First of all, let us introduce the notation: 
\beq
{\cal S}_l\,\equiv\,\Phi_{l,l}^{l,l}\,,
\label{ssiete}
\eeq
and let the ${\cal S}_l$ fields be normalized in such a way that
their two-point function is given by:
\beq
\langle\,\,{\cal S}_{l_1}\,(z_1, \bar z_1)\,\,
{\cal S}_{l_2}^{\,\dagger}\,(z_2, \bar z_2)\,\,\rangle\,=\,
\delta_{l_1 ,l_2}\,\,|\,z_1\,-\,z_2\,|^{-4h_{l_1}}\,.
\label{socho}
\eeq
Then, on general grounds, we can write the non-vanishing
correlator involving three operators of the type (\ref{ssiete})
as:
\bear
\langle\,\,{\cal S}_{l_1}\,(z_1, \bar z_1)\,\,
{\cal S}_{l_2}\,(z_2, \bar z_2)\,\,
{\cal S}_{l_1+l_2}^{\,\dagger}\,(z_3, \bar z_3)\,\,\rangle\,&=&\,
{\cal C}_{l_1, l_2}\,\,
|\,z_1\,-\,z_2\,|^{2(\,h_{l_1+l_2}-h_{l_1}-h_{l_2}\,)}\,\,\times\rc\rc
&\times&\,|\,z_1\,-\,z_3\,|^{2(\,h_{l_2}-h_{l_1}-h_{l_1+l_2}\,)}\,\,
|\,z_2\,-\,z_3\,|^{2(\,h_{l_1}-h_{l_2}-h_{l_1+l_2}\,)}\,,
\rc\rc
\label{snueve}
\eear
where ${\cal C}_{l_1, l_2}$ are constants whose value can be
obtained from the three-point functions of the 
osp$(1\vert 2)$ CFT. By using  the results of ref. \cite{osp}, one
gets:
\beq
\bigl[\,{\cal C}_{l_1, l_2}\,\bigr]^2\,\,=
\,{\Gamma \Bigl (\,{k+2\over 2k+3}\,\Bigr)
\over \Gamma \Bigl (\,{k+1\over 2k+3}\,\Bigr)}\,\,\,\,
{\Gamma \Bigl (\,{k+2+l_1+l_2\over 2k+3}\,\Bigr)\,\,
\Gamma \Bigl (\,{k+1-l_1\over 2k+3}\,\Bigr)\,\,
\Gamma \Bigl (\,{k+1-l_2\over 2k+3}\,\Bigr)\over
\Gamma \Bigl (\,{k+1-l_1-l_2\over 2k+3}\,\Bigr)\,\,
\Gamma \Bigl (\,{k+2+l_1\over 2k+3}\,\Bigr)\,\,
\Gamma \Bigl (\,{k+2+l_2\over 2k+3}\,\Bigr)}\,\,.
\label{setenta}
\eeq

It is possible to realize the parafermionic algebra in terms of
three scalar fields. This realization can be found by
representing the $\eta$ and $\xi$ fields by means of a new scalar
field $\varphi_3$:
\beq
\eta\,=\,e^{i\varphi_3}\,,
\,\,\,\,\,\,\,\,\,\,\,\,\,\,\,\,\,\,\,\,\,\,
\xi\,=\,e^{-i\varphi_3}\,.
\label{stuno}
\eeq
It can be readily proved that $\eta$ and $\xi$, as given by eq. 
(\ref{stuno}), do indeed satisfy the OPE (\ref{ciocho}) if 
$\varphi_3$ is normalized in such a way that eq. (\ref{cisiete})
also holds for $i,j=3$. Moreover, the energy-momentum tensor is:

\beq
T\,=\,-{1\over 2}\,(\partial\varphi_1)^2\,
-{1\over 2}\,(\partial\varphi_2)^2\,
-{1\over 2}\,(\partial\varphi_3)^2\,-\, 
{i\over 2\sqrt{2k+3}}\,\partial^2\,\varphi_2\,-\,
{i\over 2}\,\partial^2\,\varphi_3\,.
\label{stdos}
\eeq

Making use of eq. (\ref{stuno}), one can compute the normal
ordered products involving the $\eta$ and $\xi$ fields
which appear in the expression (\ref{suno}) of the
parafermionic currents. After some algebra one finds the form of
$\psi_{\pm 1}$ and $\psi_{\pm {1\over 2}}$ as a function of 
$\varphi_1$, $\varphi_2$ and $\varphi_3$:
\bear
\psi_1\,&=&\,\Bigl[\,
{1\over \sqrt 2}\,\partial\varphi_1\,+\,{i\over 2}\,
\sqrt{{2k+3\over k}}\,\partial\varphi_2\,+\,
{i\over 2 \sqrt k}\,\partial\varphi_3\,\Bigr]\,
e^{\sqrt{{2\over k}}\varphi_1}\,,\rc\rc
\psi_{{1\over 2}}\,&=&\,
\Bigl[\,
{1\over 2}\,\partial\varphi_1\,+\,{i\over 2}\,
\sqrt{{2k+3\over 2k}}\,\partial\varphi_2\,+\,
{i\over 2\sqrt {2k}}\,\partial\varphi_3\,
\Bigr]\,
e^{{1\over\sqrt{2k}}\varphi_1\,
-\,i\varphi_3}\,+\,
\,{1\over \sqrt{2k}}\,
e^{{1\over\sqrt{2k}}\varphi_1\,
+\,i\varphi_3}\,,\rc\rc
\psi_{-{1\over 2}}\,&=&\,
\Bigl[\,
-{1\over 2}\,\partial\varphi_1\,+\,{i\over 2}\,
\sqrt{{2k+3\over 2k}}\,\partial\varphi_2\,
-{i\over 2}\,{4k+3\over
\sqrt{2k}}\,\partial\varphi_3\,
\Bigr]\,
e^{-{1\over\sqrt{2k}}\varphi_1\,
-\,i\varphi_3}\,-\,
{1\over \sqrt{2k}}\,e^{-{1\over\sqrt{2k}}\varphi_1\,
+\,i\varphi_3}\,,\rc\rc
\psi_{-1}\,&=&\,
\Bigl[\,
-{1\over \sqrt 2}\,\partial\varphi_1\,+\,{i\over 2}\,
\sqrt{{2k+3\over k}}\,\partial\varphi_2\,
+\,{i\over 2\sqrt k}\,\partial\varphi_3\,
\Bigr]\,
e^{-\sqrt{{2\over k}}\varphi_1}\,-\,
{k+1\over \sqrt{k}}\,e^{-\sqrt{{2\over
k}}\varphi_1-2i\varphi_3}\,.\rc
\label{sttres}
\eear

Eq. (\ref{sttres}) will be particularly useful to obtain the
expression of the $\psi_l$'s for $|l|\,>\,1$ and, as a
consequence, to find the general form of the structure constants 
$C_{l, l'}$, which characterize the graded parafermionic algebra 
(\ref{seis}). The determination of these constants will be the
subject of section 5.

\medskip
\setcounter{equation}{0}
\section{Structure Constants}
\medskip
Let us, first of all, introduce some machinery which we shall
need later on in this section.  We shall work in the fully
bosonized representation of eq. (\ref{sttres}). In order to deal
with compact expressions for the currents, we shall adopt a
vector notation 
$\vec\varphi\,=\,(\,\varphi_1, \varphi_2, \varphi_3\,)$ for the
fields. The following three-dimensional vectors:
\bear
&\vec A\,=\,(\,\sqrt{{k\over 2}}\,,
\,{i\over 2}\,\sqrt{2k+3}\,,
\,{i\over 2}\,)\,,
\,\,\,\,\,\,\,\,\,\,\,\,\,\,\,\,\,\,\,
\vec B\,=\,(-\sqrt{{k\over 2}}\,,
\,{i\over 2}\,\sqrt{2k+3}\,,
\,{i\over 2}\,)\,,\rc\rc
&\vec a\,=\,(\,\sqrt{{2\over k}}\,,\,0\,,\,0\,)\,,
\,\,\,\,\,\,\,\,\,\,\,\,\,\,\,\,\,\,\,
\,\,\,\,\,\,\,\,\,\,\,\,\,\,\,\,\,\,\,
\,\,\,\,\,\,\,\,\,
\vec b\,=\,(\,0\,,\,0\,,\,2i\,)\,,
\label{stcuatro}\\
\nonumber
\eear
will become relevant in our discussion. The components of 
$\vec \varphi$ along the vectors (\ref{stcuatro}) will appear in
our representation of the parafermionic currents.

The so-called Fa\'a di Bruno (FdB)  polynomials \cite{FdB} are
defined as:
\beq
P_l\,(\,\partial f\,)\,=\,e^{-f}\,\partial^{\,l}(\,e^{f}\,)\,,
\label{stcinco}
\eeq
where $f$ is a function and $l$ is a positive integer. Notice that
the $P_l\,(\,\partial f\,)$'s are polynomials in the derivatives
of $f$. The first FdB's can
be easily obtained from (\ref{stcinco}):
\beq
P_0\,=\,1\,,
\,\,\,\,\,\,\,\,\,\,\,\,\,\,\,\,\,\,\,
P_1\,=\,\partial f\,,
\,\,\,\,\,\,\,\,\,\,\,\,\,\,\,\,\,\,\,
P_2\,=\,(\,\partial f\,)^2\,+\,\partial^2\,f\,.
\label{stseis}
\eeq
In general, these polynomials can be computed iteratively by
using  the relation:
\beq
P_{l+1}(\partial f)\,=\,\partial f\,P_{l}(\partial f)\,+\,
\partial\, \Bigl(\,P_{l}(\partial f)\,\Bigr)\,,
\label{stsiete}
\eeq
which follows easily from eq. (\ref{stcinco}). The FdB
polynomials play an important r\^ ole in the theory of integrable
hierarchies \cite{Dickey}. Their relevance in the free field
representation of the parafermionic algebra was first pointed out
in ref. \cite{Bilal}. In our case, we are going to argue that the
general expression of the positively charged currents $\psi_l$ and 
$\psi_{l+{1\over 2}}$  with $l\in\ZZ$ and $l\ge 0$ is given by:
\bear
\psi_l\,&=&\,{1\over N_l}\,P_l\,
(\,\vec A\cdot\partial\vec\varphi\,)\,\,
e^{\,l\vec a\cdot\vec\varphi}\,,\rc\rc
\psi_{l+{1\over 2}}\,&=&\,{1\over N_{l+{1\over 2}}}\,
\Bigl[\,P_{l+1}\,
(\,\vec A\cdot\partial\vec\varphi\,)\,\,
e^{\,{1\over 2}\,[\,(2l+1)\vec a\,-\,\vec b\,]\,
\cdot\,\vec\varphi}\,+\,
\,P_{l}\,
(\,\vec A\cdot\partial\vec\varphi\,)\,\,
e^{\,{1\over 2}\,[\,(2l+1)\vec a\,+\,\vec b\,]\,
\cdot\,\vec\varphi}\,\Bigr]\,,\rc\rc
&&\,\,\,\,\,\,\,\,\,\,\,\,\,\,\,\,\,\,\,\,
\,\,\,\,\,\,\,\,\,\,\,\,\,\,\,\,\,\,\,\,
\,\,\,\,\,\,\,\,\,\,\,\,\,\,\,\,\,\,\,\,
(\,l\in\ZZ\,\,,\,\,l\ge 0\,)
\label{stocho}
\eear
where $N_l$ and $N_{l+{1\over 2}}$ are normalization
constants to be determined. We shall choose these constants 
to be positive. 
Indeed, the representation given in
eq. (\ref{sttres}) for $\psi_{{1\over 2}}$ and $\psi_1$ 
coincides with the one in eq. (\ref{stocho})
if we take $N_{1/2}\,=\,\sqrt{2k}$ and $N_{1}\,=\,\sqrt{k}$. In
general, the form of the ansatz (\ref{stocho}) can be
inductively obtained by calculating the OPEs of the currents given
in (\ref{stocho}) with $\psi_1$ . Moreover, it is an easy
exercise to verify that the dimensions of the operators 
(\ref{stocho}) agree with the ones  given in eq. (\ref{diez}).

The OPEs involving  FdB
polynomials and exponentials can be computed in a
very systematic way. In appendix A we have detailed the general
procedure that one must follow. In particular, the method of
appendix A applied to the OPE of two positively charged
parafermionic currents, represented as in eq. (\ref{stocho}),
leads to the result:
\bear
\psi_l(z)\,\psi_{l'}(w)\,&=&
\,(\,1\,+\,\epsilon(l)\,\epsilon(l')\,)\,
{N_{l+l'}\over N_{l}\,N_{l'}}\,\,
{\psi_{l+l'}(w)\over (z-w)^{{2ll'\over k}\,+\,
\epsilon(l)\,\epsilon(l')}}\,+\,\cdots\,,\rc\rc
&&\,\,\,\,\,\,\,\,\,\,\,\,\,\,\,\,\,\,\,\,
\,\,\,\,\,\,\,\,\,\,\,\,\,\,\,\,\,\,\,\,
\,\,\,\,\,\,\,\,\,\,\,\,\,\,\,\,\,\,\,\,
(\,l\,,\,\,l'\ge 0\,)
\label{stnueve}
\eear
where the function $\epsilon(l)$ has been defined in eq. 
(\ref{uno}). Eq. (\ref{stnueve}) does indeed confirm the
correctness of our ansatz (\ref{stocho}). Moreover, from eq. 
(\ref{stnueve}) we obtain the value of the structure constants
for the products of two currents with positive charge in terms of
the (up to now) unknown normalization constants $N_l$, namely:
\beq
C_{l,l'}\,=\,(\,1\,+\,\epsilon(l)\,\epsilon(l')\,)\,
{N_{l+l'}\over N_{l}\,N_{l'}}\,\,,
\,\,\,\,\,\,\,\,\,\,\,\,\,\,
(\,l\,\,,\,\, l'\ge 0\,)\,.
\label{ochenta}
\eeq
In order to determine the $C_{l,l'}$'s (at least for $l, l'\ge
0$), we must first know  the $N_{l}$'s. These normalization
constants are fixed by requiring the condition written down in
eq. (\ref{catorce}), which involves the currents with negative
charge. In principle, the form of the $\psi_{-l}$'s for $l>0$ can
be determined from the expression of $\psi_{-{1\over 2}}$ and
$\psi_{-1}$ (see eq. (\ref{sttres})). It turns out, however, that
the expressions found are increasingly more complicated and it is
far from obvious to get their general form. Let us illustrate this
point with the first examples:

\bear
\psi_{-{1\over 2}}\,&=&\,{1\over N_{{1\over 2}}}\,
\Bigl[\,[\,\vec B\,-\,(k+1)\,\vec b\,]\,\cdot\,
\partial\vec\varphi\,\,
e^{\,-{1\over 2}\,(\,\vec a\,+\,\vec b\,)
\cdot\,\vec\varphi}\,\,-\,
e^{\,-{1\over 2}\,(\,\vec a\,-\,\vec b\,)\,
\cdot\,\vec\varphi}\,\Bigr]\,,\rc\rc
\psi_{-1}\,&=&\,{1\over N_{1}}\,
\Bigl[\,\,\vec B\,\cdot\,
\partial\vec\varphi\,\,
e^{\,-\vec a \cdot\,\vec\varphi}\,\,-\,
(k+1)\,e^{\,-(\vec a \,+\,\vec b\,)\,\cdot\,\vec\varphi}
\,\Bigr]\,,\rc\rc
\psi_{-{3\over 2}}\,&=&\,{1\over N_{{3\over 2}}}\,
\Bigl[\,[\,\vec B\,\cdot\partial\vec\varphi\,
\vec C\,\cdot\partial\vec\varphi\,+\,
\vec B\,\cdot\partial^2\vec\varphi\,-\,
{1\over 2}\,(k+1)\,\vec
b\,\cdot\partial^2\vec\varphi\,+\rc\rc
&+&\, {1\over 2}\,(k+1)\,\vec b\,\cdot\partial\vec\varphi\,
\vec b\,\cdot\partial\vec\varphi\,]\,
e^{\,-{1\over 2}\,(\,3\vec a\,+\,\vec b\,)\,
\cdot\,\vec\varphi}\,-\,
\vec B\,\cdot\partial\vec\varphi\,
e^{\,-{1\over 2}\,(\,3\vec a\,-\,\vec b\,)\,
\,\cdot\,\vec\varphi}
\,\,\,\Bigr]\,,\rc\rc
\psi_{-2}\,&=&\,{1\over N_{2}}\,
\Bigl[\,[\,\vec B\,\cdot\partial\vec\varphi\,
\vec B\,\cdot\partial\vec\varphi\,+\,
\vec B\,\cdot\partial^2\vec\varphi\,]\,
e^{\,-2\vec a\,\cdot\,\vec\varphi}\,-\,
2(k+1)\,(\,\vec B\,-\vec b)\cdot\partial\vec\varphi\,
e^{\,-(2\vec a\,+\,\vec b\,)\cdot\,\vec\varphi}
\,\,\,\Bigr]\,.\rc
\label{ouno}\\
\nonumber
\eear
In eq.  (\ref{ouno}) we have rewritten  $\psi_{-{1\over 2}}$ and
$\psi_{-1}$ in the more compact notation which makes use of the
vectors (\ref{stcuatro}). Of course, the different complexity  of
the parafermionic currents with positive and negative charge is
an artifact of our particular free field realization. Indeed,
one could represent the osp$(1\vert 2)$ currents in such a way
that the r\^ oles of the positive and negative charges are
interchanged and, as a consequence, the negatively charged
parafermions would have a simple general form of the type given in
eq. (\ref{stocho}). The problem here is that, in order to normalize
the currents as in eq.  (\ref{catorce}), we need to know the form
of both types of currents. Fortunately, only some part of
$\psi_{-l}$ contributes to the leading term of the
normalization OPE. 
When $l$ is a positive integer, this part is easy to
characterize. Indeed, the leading contribution to the right-hand
side of eq. (\ref{catorce}) is a c-number and, therefore, the only
terms of the negatively charged currents that are relevant are
the ones that have an exponential factor of the type $exp[-l\,\vec
a\cdot\vec\varphi\,]$ (see eq.  (\ref{stocho})). It is easy to see
that  for $l\in \ZZ$ this type of term always appears in
$\psi_{-l}$ and has the form:
\beq
\psi_{-l}\,=\,{1\over N_{l}}\,
P_l(\,\vec B\cdot\partial\vec\varphi\,)\,
e^{\,-l\vec a\,\cdot\,\vec\varphi}\,+\,\cdots\,\,,
\,\,\,\,\,\,\,\,\,\,\,\,\,\,\,\,\,
(\,l\in\ZZ \,\,,\,\,l\ge 0\,)\,.
\label{odos}
\eeq
The presence of the term (\ref{odos}) in $\psi_{-l}$ can be
verified in the expressions of $\psi_{-1}$ and $\psi_{-2}$ written
in eq.  (\ref{ouno}) and it is not hard to prove it in general. By
applying the methods of appendix A, we can now compute the leading
term of the OPE $\psi_l(z)\,\psi_{l'}(w)$ and determine the value
of
$N_l$ for $l\in\ZZ$. The result one arrives at is rather simple:
\beq
(\,N_{l}\,)^2\,=\,{k!\,l!\over (k-l)!}\,\,,
\,\,\,\,\,\,\,\,\,\,\,\,\,\,\,\,\,
(\,l\in\ZZ \,\,,\,\,l\ge 0\,)\,.
\label{otres}
\eeq

If we knew the normalization constants $N_{l+{1\over 2}}$ for the
half-integer currents, we would be able to write the general form
of the $C_{l,l'}$ constants for $l,l'>0$. However, the direct
calculation of $N_{l+{1\over 2}}$ is very difficult due to our
failure in finding the general form of the term in 
$\psi_{-l-{1\over 2}}$ that contributes to its normalization. In
the case of the structure constants $C_{l,l'}$ that involve both
positive and negative charges, the situation is even worse and
only some particular cases can be computed directly. Some of
these constants are:
\bear
C_{l,-{1\over 2}}\,&=&\,-l\,\,{N_{l-{1\over 2}}
\over N_l\,N_{{1\over 2}}}\,,
\,\,\,\,\,\,\,\,\,\,\,\,\,\,\,\,\,
\,\,\,\,\,\,\,\,\,\,\,\,\,\,\,\,\,
\,\,\,\,\,\,\,\,\,\,\,\,\,\,\,\,\,\,\,\,\,
C_{l+{1\over 2},-{1\over 2}}\,=\,2(\,k-l\,)\,\,{N_{l}
\over N_{l+{1\over 2}}\,N_{{1\over 2}}}\,,\rc\rc\rc
C_{l,-1}\,&=&\,l(\,k-l+1\,)\,\,{N_{l-1}
\over N_l\,N_{1}}\,,
\,\,\,\,\,\,\,\,\,\,\,\,\,\,\,\,\,
\,\,\,\,\,\,\,\,\,\,\,\,\,\,\,\,\,
C_{l+{1\over 2},-1}\,=\,l(\,k-l\,)\,\,{N_{l-{1\over 2}}
\over N_{l+{1\over 2}}\,N_{1}}\,.
\rc\rc
&&\,\,\,\,\,\,\,\,\,\,\,\,\,\,\,\,\,
\,\,\,\,\,\,\,\,\,\,\,\,\,\,\,\,\,
\,\,\,\,\,\,\,\,\,\,\,\,\,\,\,\,\,
(\,l\in \ZZ\,\,,\,\,l\ge 0\,)
\label{ocuatro}
\eear

Remarkably, the knowledge of the explicit results
(\ref{ochenta}), (\ref{otres})  and (\ref{ocuatro})  is enough to
obtain the full set of structure constants. Let us explain how
this can be done by using the associativity condition of the
parafermionic algebra. Let us consider the 
three-point correlator 
$\langle\,\psi_{m_1}(z_1)\,\psi_{m_2}(z_2)
\,\psi_{m_3}(z_3)\,\rangle$ which, due to charge conservation, is
only non-vanishing if  $m_1+m_2+m_3\,=\,0$. The dependence of
this correlator on the coordinates is fixed by conformal
invariance. This coordinate dependence can be compared with the
one that is obtained by making use of the parafermionic OPEs 
(\ref{seis}). These OPEs can be computed in two different forms, 
that differ in the way in which we associate the currents. By
comparing these two results with the exact value of the
correlator, we obtain the following condition for the structure
constants:
\beq
C_{m_1, m_2+m_3}\,C_{m_2, m_3}\,=\,
C_{m_1, m_2}\,C_{m_1+m_2, m_3}\,\,,
\,\,\,\,\,\,\,\,\,\,\,\,\,\,\,\,\,\,\,\,\,\,\,
(\,m_1+m_2+m_3\,=\,0\,)\,.
\label{ocinco}
\eeq

Let us consider now eq. (\ref{ocinco}) for 
$m_1\,=\,l_2-l_1$, $m_2\,=\,l_1$ and
$m_3\,=\,-l_2$ with $l_1, l_2\ge 0$ and $l_2\ge l_1$. In this
case, eq. (\ref{ocinco}) reduces to:
\beq
C_{l_1, -l_2}\,=\,C_{l_2-l_1, l_1}\,\,,
\,\,\,\,\,\,\,\,\,\,\,\,\,\,\,\,\,\,
(\,l_2\ge l_1\ge 0\,)\,.
\label{oseis}
\eeq
In order to obtain eq. (\ref{oseis}) we have taken into account
that, according to eqs. (\ref{catorce}) and (\ref{quince}), one
has:  
\beq
C_{l, -l}\,=\,(-1)^{\epsilon(l)}\,C_{-l,l}\,=\,1\,\,,
\,\,\,\,\,\,\,\,\,\,\,\,\,\,\,\,\,\,
(\,l\ge 0\,)\,.
\label{osiete}
\eeq

Notice that eq. (\ref{oseis}) relates some structure constants
involving positive and negative charges to the $C_{l_1, l_2}$'s
for $l_1,l_2\ge 0$. In particular, as $C_{l_1, l_2}\,\ge\,0$ for
$l_1,l_2\ge 0$ (see eq.~(\ref{ochenta}) and recall that the
normalization constants are always positive), eq. (\ref{oseis})
implies that $C_{l_1, -l_2}\,\ge\,0$ for $l_2\ge l_1\ge 0$.  

The condition (\ref{ocinco}) can be used to determine the
behaviour of the structure constants $C_{l_1, l_2}$ when the
signs of both charges $l_1$ and $l_2$ are reversed. In general,
due to the equivalence between the positive and negative charge
sectors of the theory, $C_{l_1, l_2}$ and $C_{-l_1, -l_2}$ could
differ at most by a sign. Accordingly, let us put:
\beq
C_{-l_1, -l_2}\,=\,\sigma(l_1,l_2)\,C_{l_1,l_2}\,,
\label{oocho}
\eeq
where $\sigma(l_1,l_2)$ is a sign which  can be
determined by using the associativity condition. Indeed, let us
take in  eq. (\ref{ocinco}) 
$m_1\,=\,-l_1$,  $m_2\,=\,-l_2$ and $m_3\,=\,l_1+l_2$,
with $l_1, l_2\ge 0$. Using again eq. (\ref{osiete}), we get:
\beq
C_{-l_1, -l_2}\,=\,(-1)^{\epsilon (l_2)}\, 
C_{-l_2, l_1+l_2}\,\,,
\,\,\,\,\,\,\,\,\,\,\,\,\,\,\,\,\,\,
(\,l_1\,,\,l_2\,\ge 0\,)\,,
\label{onueve}
\eeq
which, after taking eqs. (\ref{oseis}) and (\ref{oocho}) into
account, reduces to the following equation for $\sigma(l_1,l_2)$:
\beq
\sigma(l_1,l_2)\,=\,(-1)^{\epsilon (l_2)}
\sigma(l_2,-l_1-l_2)\,.
\label{noventa}
\eeq
In order to solve eq.  (\ref{noventa}), we assume that
$\sigma(l_1,l_2)$ only depends  on  the integer or
half-integer nature of $l_1$ and $l_2$ and is a symmetric
function of its arguments. Adopting the ansatz:
\beq
\sigma(l_1,l_2)\,=\,
(-1)^{\alpha(\epsilon (l_1)+\epsilon (l_2))\,+\,
\beta \epsilon (l_1)\epsilon (l_2) }\,,
\label{nuno}
\eeq
one easily gets that eq. (\ref{noventa}) is solved if
$\beta\,=\,\alpha-1$. This solution can be completely determined
by looking at some particular cases of $l_1$ and $l_2$. So, for
example, it is clear from eq. (\ref{vuno}) that 
$\sigma(\,{1\over 2},{1\over 2}\,)\,=-1$ and 
$\sigma(\,1,-{1\over 2}\,)\,=\,+1$. These values are reproduced
by our solution if $\alpha\,=\,0$. Thus, one has:
\beq
C_{-l_1, -l_2}\,=\,
(-1)^{\epsilon (l_1)\epsilon (l_2) }
C_{l_1,l_2}\,.
\label{ndos}
\eeq

The behaviour of eq. (\ref{ndos}) can be checked in some other
particular cases in which the structure constants can be  computed
directly.
This verification gives support to the hypothesis we have assumed
in the derivation of eq. (\ref{ndos}). On the other hand, notice
that eq. (\ref{ndos}) gives the structure constants involving two
negative charges in terms of the constants for two positive
charges. It is also possible to relate the constants $C_{l_1,
-l_2}$ for 
$l_1\,>l_2\,\ge\, 0$ (\ie\ the case not included in eq.
(\ref{oseis})) to the constants given in eq.~(\ref{ochenta}).
This can be achieved by combining our previous relations. First
of all, if $l_1\,>l_2\,\ge\, 0$, eq. (\ref{quince}) allows one to
write:
\beq
C_{l_1, -l_2}\,=\,(-1)^{\epsilon(l_2)}\,C_{-l_2, l_1}\,\,,
\,\,\,\,\,\,\,\,\,\,\,\,\,\,\,\,\,\,
(\,l_1\,>l_2\,\ge\, 0\,)\,.
\label{ntres}
\eeq
Moreover, by using eq. (\ref{ndos}) on the right-hand side of eq.
(\ref{ntres}), one arrives at:
\beq
C_{l_1, -l_2}\,=\,
(-1)^{\epsilon(l_2)\,(\,\epsilon(l_1)\,+\,1\,)}\,
C_{l_2, -l_1}\,\,,
\,\,\,\,\,\,\,\,\,\,\,\,\,\,\,\,\,\,
(\,l_1\,>l_2\,\ge\, 0\,)\,.
\label{ncuatro}
\eeq
Finally, by employing the relation (\ref{oseis}), we can write:
\beq
C_{l_1, -l_2}\,=\,
(-1)^{\epsilon(l_2)\,(\,\epsilon(l_1)\,+\,1\,)}\,
C_{l_1-l_2, l_2}\,\,,
\,\,\,\,\,\,\,\,\,\,\,\,\,\,\,\,\,
(\,l_1> l_2\ge 0\,)\,.
\label{ncinco}
\eeq

Notice that the right-hand side of eq. (\ref{ncinco}) can be
evaluated by means of eq. (\ref{ochenta}). Taken together, eqs. 
(\ref{oseis}) and (\ref{ncinco})  give the structure constants
for the product of a current of positive charge and a current of
negative charge in terms of the $C_{l_1, l_2}$'s with $l_1,
l_2\ge 0$. Let us write this result as:
\beq
C_{l_1, -l_2}\,=\,
\cases{C_{l_2-l_1, l_1}&if $l_2\,\ge\,l_1\,\ge 0$\cr\cr
(-1)^{\epsilon(l_2)\,(\,\epsilon(l_1)\,+\,1\,)}\,
C_{l_1-l_2, l_2}&if $l_1\,>\,l_2\,\ge 0\,.$}
\label{nseis}
\eeq

Eqs. (\ref{nseis}) and (\ref{ndos}) reduce the problem of the
determination of the structure constants to the evaluation of the
right-hand side of eq. (\ref{ochenta}), \ie\ to the calculation
of the normalization constants $N_l$. Recall that we have found a
partial answer to this problem (see eq. (\ref{otres})). It
remains to compute the normalization constants corresponding to
the currents with half-integer charge. It turns out that the
relations we have found allow one to express these constants in
terms of the ones written in eq. (\ref{otres}). Let us, first of
all, substitute in eq. (\ref{oseis}) $l_1\,=\,{1\over 2}$ and
$l_2=l$ with $l>0$. Eq. (\ref{oseis}) reduces in this case to:
\beq
C_{{1\over 2}, -l}\,=\,C_{l-{1\over 2}, {1\over 2}}\,\,,
\,\,\,\,\,\,\,\,\,\,\,\,\,\,\,\,\,
(\,l>0\,\,,\,\,l\in\ZZ\,)\,.
\label{nsiete}
\eeq
On the other hand, the right-hand side of eq. (\ref{nsiete}) can
be computed  by means of eq. (\ref{ochenta}) and the result is:
\beq
C_{l-{1\over 2}, {1\over 2}}\,=\,2\,
{N_l\over N_{l-{1\over 2}}\,N_{1\over 2}}\,\,,
\,\,\,\,\,\,\,\,\,\,\,\,\,\,\,\,\,
(\,l>0\,\,,\,\,l\in\ZZ\,)\,,
\label{nocho}
\eeq
whereas, after using eqs. (\ref{quince}) and (\ref{ndos}), the
constant appearing on the left-hand side of eq. (\ref{nsiete})
can be reduced to one of the cases of eq. (\ref{ocuatro}), namely:
\beq
C_{{1\over 2}, -l}\,=\,-C_{l,-{1\over 2}}\,=\,
l\,{N_{l-{1\over 2}}\over N_l\,N_{{1\over 2}}}\,\,,
\,\,\,\,\,\,\,\,\,\,\,\,\,\,\,\,\,
(\,l>0\,\,,\,\,l\in\ZZ\,)\,.
\label{nnueve}
\eeq
By substituting eqs. (\ref{nocho}) and (\ref{nnueve}) in eq. 
(\ref{nsiete}), we find:
\beq
(\,N_{l-{1\over 2}}\,)^2\,=\,{2\over l}\,
(\,N_{l}\,)^2\,\,,
\,\,\,\,\,\,\,\,\,\,\,\,\,\,\,
(\,l\in\ZZ\,,\,l>0\,)\,,
\label{cien}
\eeq
which is the announced relation. From eqs. (\ref{otres}) and
(\ref{cien}),  we can write the general expression of $N_l$ 
for arbitrary $l$:
\beq
(\,N_{l}\,)^2\,=\,(\,1\,+\,\epsilon(l)\,)\,
{k!\,[\,l\,]!\over [\,k-l\,]!}\,.
\label{ctuno}
\eeq
In eq. (\ref{ctuno}) the brackets denote integer part (as in eq. 
(\ref{dos})). It is now immediate to find the structure
constants. Indeed, after substituting eq. (\ref{ctuno}) on the
right-hand side of eq.~(\ref{ochenta}), we find the 
simple expression:
\beq
C_{l,l'}^2\,=\,
{[\,l\,+\,l'\,]!\,[\,k\,-\,l\,]!\,[\,k\,-\,l'\,]!\over
k!\,[\,l\,]!\,[\,l'\,]!\,[\,k\,-\,l\,-\,l'\,]!}\,\,,
\,\,\,\,\,\,\,\,\,\,\,\,\,\,\,\,\,\,\,\,\,\,\,\,\,\,\,\,\,\,
(\,l\,,\,l'\ge 0\,)\,.
\label{ctdos}
\eeq

Several remarks concerning eq. (\ref{ctdos}) are in order. First
of all, it is worthwhile to point out that we must take the
positive sign of the square root when computing $C_{l,l'}$ from 
eq.~(\ref{ctdos})  (recall that $C_{l,l'}\,\ge\,0$ for
$l,l'\,\ge\,0$). Secondly, although eqs. (\ref{ctdos}), 
(\ref{nseis}) and (\ref{ndos}) have been obtained in an indirect
way by means of a chain or arguments based on eq. (\ref{ocinco}),
their predictions can be successfully compared with the results
of direct calculations such as the ones written in eq.
(\ref{ocuatro}). 
Moreover, it is interesting to stress the
differences and similarities between our result and the one
corresponding to the $Z_k$ parafermions. It is clear from the
comparison of eqs. (\ref{ctdos}), (\ref{nseis}) and (\ref{ndos})
and the values given in ref. \cite{ZF} for the structure constants
that, when the charges are integer, our results coincide with
those of ref.~\cite{ZF}. Indeed, if $l, l'\,\in\ZZ$,  one can
eliminate the integer part symbol from the right-hand side of eq.
(\ref{ctdos})  and, on the other hand, all minus signs in eqs.
(\ref{ndos}) and  (\ref{nseis}) and (\ref{quince}) disappear. 
On the contrary, 
when any of the charges is half-integer, our eq. (\ref{ctdos})
differs from the result of ref. \cite{ZF} and  minus signs, which
reflect the graded nature of our system, do appear in some of the
structure constants.

\medskip
\setcounter{equation}{0}
\section{Concluding Remarks}
\medskip

In this paper we have formulated a graded generalization of the
parafermionic symmetry. By means of simple first-principle
arguments we have been able to determine the general form of the
algebra and, in particular, the conformal dimensions of the
parafermionic currents. The results of this general analysis
allowed us to identify the graded parafermionic system with the 
osp$(1\vert 2)/U(1)$ CFT. By using this identification we have
been able to prove the existence of a graded extension of the
parafermionic symmetry and, actually, we have found a free field
realization. The central charge $c$ of the model can be easily
obtained from its osp$(1\vert 2)/U(1)$ representation. The fact
that $c<0$ implies that the theory cannot be unitary.

The parafermionic Hilbert space can be represented as a direct sum
of highest-weight modules. The modes of the currents satisfy
generalized (anti)commutation relations on this Hilbert space
which determine the conformal dimensions of the operators of the
field space of the model. The free field representation of our
system can be used to obtain the structure constants of the
current algebra, a highly non-trivial result which is a
generalization of the one in ref. \cite{ZF}.

The ordinary parafermions were introduced in ref. \cite{ZF} as a
generalization of the Ising model and, in general, they describe
self-dual critical points in $Z_k$ statistical systems. It would
be desirable to have a similar  interpretation for the CFT
constructed in this paper. Notice that, although our model is very
similar to the $Z_k$ parafermionic system, there exist
substantial differences between them. First of all, there is the
unitarity issue.  Secondly, the set of conformal weights written
in eq. (\ref{cicuatro}) and those corresponding to the $Z_k$
parafermions are very different as a consequence of the different
definitions of the highest-weight modules. Indeed, in our case,
the highest-weight conditions (\ref{ccuatro}) involve both the
bosonic $A^{(\pm)}$ and the fermionic $B^{(\pm)}$ mode operators,
whereas, on the contrary, only the $A^{(\pm)}$'s are used to
define the highest-weight primary fields of the $Z_k$
parafermionic theory. Despite  these difficulties, our system
displays many good properties from the  representation theory
point of view and, for this reason, we believe that the symmetry
studied could be important in the characterization of some
critical statistical mechanics models.

There are some other interesting aspects of the graded
parafermionic system which were not considered here. Let us
mention some of them. First of all, we could compute the
characters of the theory which, according to its representation
as an osp$(1\vert 2)/U(1)$ coset, are nothing but string functions
of the osp$(1\vert 2)$ affine superalgebra \cite{KP}. In the
framework of the free field realization we could, in principle,
develop a BRST formalism for the study of these characters,
similar to the one constructed in ref. \cite{Distler} for the
$su(2)$ case. Moreover, one could analyze more general coset
constructions involving the  osp$(1\vert 2)$  superalgebra. The
parafermionic theory we have  studied in this paper is just a
particular example of these coset  theories. However, in analogy
with what happens in the $su(2)$  case, it might be that the
graded parafermions are the building blocks of the
osp$(1\vert 2)$ coset models. Finally, it would be important to
find out if the osp$(1\vert 2)$ parafermions admit integrable
deformations, similar to the ones  that the $su(2)$ theory has
\cite{Fateev}. In case of affirmative answer we would find new
families of massive integrable two-dimensional theories.

\section{ Acknowledgments}
We are grateful to I. P. Ennes and J. L. Miramontes for useful
discussions and a critical reading of the manuscript. This work was
supported in part by DGICYT under grant PB96-0960,  by CICYT under
grant  AEN96-1673 and by the European Union TMR grant
ERBFMRXCT960012.

\newpage

\vskip 1cm                                               
{\Large{\bf APPENDIX A}}                                 
\vskip .5cm                                              
\renewcommand{\theequation}{\rm{A}.\arabic{equation}}  
\setcounter{equation}{0}  

In this appendix we are going to develop a method which allows
the computation of OPEs of FdB polynomials and exponentials in a
rather systematic way. Let us consider two operators  
$O_1(\,z\,)$ and $O_2(\,w\,)$, which have the generic form:
\bear
O_1(\,z\,)\,&=&\,P_{l_1}\,(\,\vec A_1\,\cdot\,\partial \vec\varphi\,(\,z\,)\,)\,
e^{\vec a_1\,\cdot\,\vec \varphi\,(\,z\,)}\,,
\rc\cr
O_2(\,w\,)\,&=&\,P_{l_2}\,(\,\vec A_2\,\cdot\,\partial \vec\varphi\,(\,z\,)\,)\,
e^{\vec a_2\,\cdot\,\vec \varphi\,(\,z\,)}\,,\rc
\label{apauno}
\eear
where $P_{l_1}$ and $P_{l_2}$ are FdB polynomials (defined in
eq. (\ref{stcinco})) and $\vec A_1$, $\vec A_2$, $\vec a_1$ and 
$\vec a_2$ are constant vectors. Our aim is to give an
expression of the OPE $O_1(\,z\,)\,O_2(\,w\,)$. Recall (see eq. 
(\ref{stcinco})) that the FdB polynomials are defined through the
derivative of an exponential. Let us start our calculation by
carrying out a point-splitting procedure, which distinguishes the
coordinates of the fields where the derivative acts from the
others. Therefore, instead of the expressions (\ref{apauno}), we
shall consider the following bilocal representation of $O_1$ and
$O_2$:
\bear
O_1(\,z\,)\,&=&\,\partial_{z_2}^{l_1}\,\Bigl [\,
e^{(\,\vec a_1\,-\,\vec A_1\,)\,\cdot\,\vec \varphi\,(\,z_1\,)\,+\,
\vec A_1\,\cdot\,\vec \varphi\,(\,z_2\,)}\,\Bigr]\,,\rc\rc
O_2(\,w\,)\,&=&\,\partial_{w_2}^{l_2}\,\Bigl [\,
e^{(\,\vec a_2\,-\,\vec A_2\,)\,\cdot\,\vec \varphi\,(\,w_1\,)\,+\,
\vec A_2\,\cdot\,\vec \varphi\,(\,w_2\,)}\,\Bigr]
\,.\label{apados}\\
\nonumber
\eear
In eq. (\ref{apados}) (and in what follows) it is implicitly
understood that after the derivatives are performed one must take
the coincidence limit $z_1\rightarrow z_2\rightarrow z$ and  
$w_1\rightarrow w_2\rightarrow w$ (although this limit will no be
indicated explicitly). The obvious advantage of eq.
(\ref{apados}) with respect to our original expressions
(\ref{apauno}) is that the  expansion of the product of two
bilocal operators (\ref{apados}) reduces to the OPE (of the
derivatives) of two exponentials. Actually, let $F(z,w)$ be the
function:

\beq
F(z,w)\,=\,(\,z_1\,-\,w_1\,)^{\alpha_{11}}\,
(\,z_1\,-\,w_2\,)^{\alpha_{12}}\,
(\,z_2\,-\,w_1\,)^{\alpha_{21}}\,
(\,z_2\,-\,w_2\,)^{\alpha_{22}}\,,
\label{apatres}
\eeq
where the $\alpha_{ij}$'s are the following scalar products:

\bear
\alpha_{11}\,&=&\,-(\,\vec a_1\,-\,\vec A_1\,)\,\cdot\,
(\,\vec a_2\,-\,\vec A_2\,)\,,
\,\,\,\,\,\,\,\,\,\,\,\,\,\,\,\,\,\,\,\,
\,\,\,\,\,\,\,\,\,\,\,\,\,\,\,\,\,\,\,\,
\alpha_{12}\,=\,-(\,\vec a_1\,-\,\vec A_1\,)\,\cdot\,
\vec A_2\,,\rc\rc
\alpha_{21}\,&=&-\vec A_1\,\cdot\,(\,\vec a_2\,-\,\vec A_2\,)\,,
\,\,\,\,\,\,\,\,\,\,\,\,\,\,\,\,\,\,\,\,
\,\,\,\,\,\,\,\,\,\,\,\,\,\,\,\,\,\,\,\,
\,\,\,\,\,\,\,\,\,\,\,\,\,\,\,\,\,\,\,\,\,\,\,
\alpha_{22}\,=\,-\vec A_1\,\cdot\,\vec A_2\,.\rc
\label{apacuatro}
\eear
By using the well-known rules for the computation of OPEs of
exponentials of scalar fields, we immediately get:
\bear
O_1(\,z\,)\,O_2(\,w\,)\,=\,\partial_{z_2}^{l_1}\,\,\partial_{w_2}^{l_2}\,
\,\Bigl[\,F(z,w)\,
e^{(\,\vec a_1\,-\,\vec A_1\,)\,\cdot\,\vec \varphi\,(\,z_1\,)\,+\,
\vec A_1\,\cdot\,\vec \varphi\,(\,z_2\,)\,+\,
(\,\vec a_2\,-\,\vec A_2\,)\,\cdot\,\vec \varphi\,(\,w_1\,)\,+\,
\vec A_2\,\cdot\,\vec \varphi\,(\,w_2\,)}\,\Bigr]\,.\rc
\label{apacinco}
\eear
In order to get the final result we must evaluate the derivatives
and then take the $z_1\rightarrow z_2\rightarrow z$ and  
$w_1\rightarrow w_2\rightarrow w$ limits. We shall use Leibnitz's
rule, which, for the $l^{{\rm th}}$ derivative of the product of
two functions $f$ and $g$, can be written as:
\beq
\partial^l\,(\,f\,g\,)\,=\,\sum_{r=0}^{l}\,
{l \choose r}\,\partial^r\,f\,\partial^{l-r}\,g\,.
\label{apaseis}
\eeq
Let us write the coincidence limit of the derivatives of the
function $F$ as:
\beq
\partial_{z_2}^{r}\,\,\partial_{w_2}^{s}\Bigl[\,F(z,w)\,\,\Bigr]\,=\,
{1\over (z-w)^{\vec a_1\,\cdot\,\vec a_2}}\,\, 
{a_{rs}\over (\,z\,-\,w\,)^{r+s}} \,,
\label{apasiete}
\eeq
where the $a_{rs}$ are c-numbers and the coordinate dependence is
fixed by the exponents $\alpha_{ij}$ of eq. (\ref{apacuatro}).
(Notice that the sum of the $\alpha_{ij}$'s is 
$-\vec a_1\,\cdot\,\vec a_2$). Making use of these results, we can
write:
\bear
O_1(\,z\,)\,O_2(\,w\,)\,&=&\,
{1\over (z-w)^{\vec a_1\,\cdot\,\vec a_2}}\,\, 
\sum_{r=0}^{l_1}\,\sum_{s=0}^{l_2}\,\,
{l_1\choose r}\,{l_2\choose s}\,
{a_{rs}\over (\,z\,-\,w\,)^{r+s}}\times\rc\rc
&\times&\,P_{l_1-r}\,(\,\vec A_1\,\cdot\,\partial \vec\varphi\,(\,z\,)\,)
\,\,P_{l_2-s}\,(\,\vec A_2\,\cdot\,\partial \vec\varphi\,(\,w\,)\,)\,
e^{\vec a_1\,\cdot\,\vec \varphi\,(\,z\,)\,+\,
\vec a_2\,\cdot\,\vec \varphi\,(\,w\,)}\,.
\rc\rc
\label{apaocho}
\eear
In order to obtain eq. (\ref{apaocho}) it is essential to
realize that the derivatives of the exponentials in eq.
(\ref{apacinco}) can be organized again as FdB polynomials. In
this way, we get on the right-hand side of eq. (\ref{apaocho})
products of operators of the same type as in eq. (\ref{apauno}).
The constants $a_{rs}$ can be obtained by evaluating explicitly
the derivatives of $F$. In view of the 
form of $F$ (eq.~(\ref{apatres})), 
this calculation reduces to the computation of
the derivatives of powers. One gets:

\bear
a_{rs}\,=\,(-1)^s\,\,\sum_{n=0}^{r}\,\,\sum_{m=0}^{s}\,\,
{r\choose n}\,{s\choose m}\,
{\Gamma(\,\alpha_{21}+1\,)\,\Gamma(\,\alpha_{12}+1\,)\,
\Gamma(\,\alpha_{22}+1\,)\over
\Gamma(\,\alpha_{21}-n+1)\,\Gamma(\,\alpha_{12}-m+1\,)\,
\Gamma(\,\alpha_{22}+m+n-r-s+1\,)}\,.\rc\rc
\label{apanueve}
\eear

Let us illustrate our method in the case of the expansion of the
product of two currents  $\psi_{l_1}(z)$ and $\psi_{l_2}(w)$,
where $l_1$ and  $l_2$ are positive integers. The free-field 
expression of the currents $\psi_{l_1}(z)$ and $\psi_{l_2}(w)$ is
given in the first equation in (\ref{stocho}). Apart from the
normalization constants $N_{l_1}$ and $N_{l_2}$,  $\psi_{l_1}(z)$
and $\psi_{l_2}(w)$ are of the form (\ref{apauno}) with:
\beq
\vec A_1\,=\,\vec A_2\,=\,\vec A\,,
\,\,\,\,\,\,\,\,\,\,\,\,\,\,\,
\vec a_1\,=\,l_1\,\vec a\,,
\,\,\,\,\,\,\,\,\,\,\,\,\,\,\,
\vec a_2\,=\,l_2\,\vec a\,,
\label{apadiez}
\eeq
where the vectors $\vec A$ and $\vec a$ have been defined in eq. 
(\ref{stcuatro}). It is straightforward to compute the value of
the scalar products $\alpha_{ij}$ in this case. The result is:
\bear
\alpha_{11}\,&=&l_1\,+\,l_2\,+1\,-\,{2l_1l_2\over k}\,,
\,\,\,\,\,\,\,\,\,\,\,\,\,\,\,\,\,\,\,\,
\,\,\,\,\,\,\,\,\,\,\,\,\,\,\,\,\,\,\,\,
\alpha_{12}\,=\,-l_1-1\,,\rc\rc
\alpha_{21}\,&=&-l_2-1\,,
\,\,\,\,\,\,\,\,\,\,\,\,\,\,\,\,\,\,\,\,
\,\,\,\,\,\,\,\,\,\,\,\,\,\,\,\,\,\,\,\,
\,\,\,\,\,\,\,\,\,\,\,\,\,\,\,\,\,\,\,\,\,\,\,\,\,\,\,\,\,\,\,\,
\alpha_{22}\,=\,1\,.
\rc
\label{apaonce}
\eear
Moreover, the sum (\ref{apanueve}), which gives the value of
$a_{rs}$, can be easily obtained since it truncates. One gets:
\beq
a_{rs}\,=\,{(-1)^r\over l_1!l_2!}\,(\,l_1+s-1\,)!\,(\,l_2+r-1\,)!\,
(\,l_1l_2-rs\,)\,.
\label{apadoce}
\eeq
Substituting eq. (\ref{apadoce}) in our general expression
(\ref{apaocho}), we get:
\bear
\psi_{l_1}(z)\,\psi_{l_2}(w)\,&=&\,{1\over N_{l_1}N_{l_2}}\,
{1\over (z-w)^{{2l_1l_2\over k}}}\,
\sum_{r=0}^{l_1}\,\sum_{s=0}^{l_2}\,\,
{l_1\choose r}\,{l_2\choose s}\,
{(-1)^r\over l_1!l_2!}\,(\,l_1+s-1\,)!\,(\,l_2+r-1\,)!\,\times\rc\rc\rc
&&\times\,{(\,l_1l_2-rs\,)\over (z-w)^{r+s}}\,\,
P_{l_1-r}\,(\,\vec A\,\cdot\,\partial \vec\varphi\,(\,z\,)\,)
\,\,P_{l_2-s}\,(\,\vec A\,\cdot\,\partial \vec\varphi\,(\,w\,)\,)\,
e^{l_1\vec a\,\cdot\,\vec \varphi\,(\,z\,)\,+\,
l_2\vec a\,\cdot\,\vec \varphi\,(\,w\,)}\,,
\rc\rc
&&\,\,\,\,\,\,\,\,\,\,\,\,\,\,\,\,\,\,\,\,\,\,\,\,\,\,\,\,\,\,\,\,
(\,l_1\,,\,l_2\,\ge 0\,\,\,\,\,l_1\,,\,l_2\,\in\ZZ\,)
\label{apatrece}
\eear

The right-hand side of eq. (\ref{apatrece}) should match the
first equation in (\ref{trece}). It is clear that, in order to
compare these two equations, we must expand in Taylor series the
fields evaluated at $z$ on the right-hand side of eq. 
(\ref{apatrece}). A priori, it is not obvious that the leading
singularity of the OPE (\ref{apatrece}) coincides 
with the one displayed in eq. (\ref{trece}). Actually, this
coincidence only takes place if all the anomalous terms,
present on the right-hand side of eq. (\ref{apatrece}), 
cancel. This cancelation is a highly non-trivial fact and is a
consequence of the following identity satisfied by the FdB
polynomials:

\bear
&&\sum_{r={\rm max} (0, p-l_2)}^{l_1}\,\,
\sum_{s={\rm max} (0, p-r)}^{l_2}\,\,
{l_1\choose r}\,{l_2\choose s}\,
{(-1)^r\over l_1!l_2!}\,(\,l_1+s-1\,)!\,(\,l_2+r-1\,)!\,\times\rc\rc\rc
&&\times\,{(\,l_1l_2-rs\,)\over (r+s-p)!}\,\,
\partial^{r+s-p}\,
P_{l_1-r}\,(\,\vec A\,\cdot\,\partial \vec\varphi\,(\,w\,)\,)
\,\,P_{l_2-s}\,(\,\vec A\,\cdot\,\partial \vec\varphi\,(\,w\,)\,)\,=\,\rc\rc\rc
&&=\,\,
\delta_{p,0}\,P_{l_1+l_2}\,
(\,\vec A\,\cdot\,\partial \vec\varphi\,(\,w\,)\,)\,,
\,\,\,\,\,\,\,\,\,\,\,\,\,\,\,\,\,\,\,\,\,\,\,\,\,\,\,\,\,\,\,\,
(\,l_1\,,\,l_2\,\ge 0\,\,\,\,\,l_1\,,\,l_2\,\in\ZZ\,)\,.
\label{apacatorce}
\eear
Eq. (\ref{apacatorce}) can be verified by direct calculation (by 
making use of the explicit form of the FdB polynomials). Notice
that the terms on the left-hand side of eq. (\ref{apacatorce}) are
precisely the ones generated when the FdB polynomials on the
right-hand side of eq. (\ref{apatrece}) are expanded in Taylor
series (\ie\ one does not need to expand the exponential in order
to cancel the anomalies). Therefore, the OPE of two currents can
be written as:
\bear
\psi_{l_1}(z)\,\psi_{l_2}(w)\,&=&\,{1\over N_{l_1}N_{l_2}}\,\,
{P_{l_1+l_2}\,(\,\vec A\,\cdot\,\partial \vec\varphi\,(\,w\,)\,)\,\,
e^{(l_1+l_2)\,\vec a\,\cdot\,\vec \varphi\,(\,w\,)\,}\over
(z-w)^{{2l_1l_2\over k}}}\,+\,\cdots\,\,=\rc\rc\rc
&&=\,{N_{l_1+l_2}\over N_{l_1}N_{l_2}}\,
\psi_{l_1+l_2}\,(w)\,+\,\cdots\,,
\rc\rc
&&\,\,\,\,\,\,\,\,\,\,\,\,\,\,\,\,\,\,\,\,\,\,\,\,\,\,\,\,\,\,\,\,
(\,l_1\,,\,l_2\,\ge 0\,\,\,\,\,l_1\,,\,l_2\,\in\ZZ\,)
\label{apaquince}
\eear
a result which agrees with that written in eq.
(\ref{stnueve}). In the case in which the charges $l_1$ and/or
$l_2$ are positive half-integers, the corresponding OPEs can be
computed in a similar way. In this case we must also use the
second equation in (\ref{stocho}). It can be checked that the
anomaly cancelation also takes place in this case, again  as a
consequence of the identity (\ref{apacatorce}). The general result
of these OPEs has been written in eq. (\ref{stnueve}).

The method just described can also be employed to compute the
normalization constants $N_l$ (see eq. (\ref{otres})) and other
structure constants (such as the ones in eq. (\ref{ocuatro})).


\begin{thebibliography}{99}

\bibitem{CFT} For a review see, for example, 
S. Ketov, {\sl ``Conformal Field Theory"}, (World Scientific,
Singapore, 1995) and P. Di Francesco, P. Mathieu and D.
S\'en\'echal,  {\sl Conformal Field Theory}, (Springer-Verlag,
New York, 1997). 


\bibitem{para} L. Kadanoff and H. Ceva, {\sl \pr} {\bf B3} (1971)
3918; E. Fradkin and  L. Kadanoff, {\sl \np} {\bf B170} (1980)1.




\bibitem{ZF}A. B. Zamolodchikov and V. A. Fateev, 
{\sl \jept} {\bf 62} (1985) 215. 


\bibitem{GKO} P. Goddard, A. Kent and D. Olive, {\sl \pl} 
{\bf B152} (1985) 88; {\sl \cmp} {\bf 103} (1986) 105.


\bibitem{Gepner}D. Gepner, {\sl \np} {\bf B290} [FS20] (1987) 10.






\bibitem{ZFdos} A. B. Zamolodchikov and V. A. Fateev, 
{\sl \jept} {\bf 63} (1986) 913; {\sl \tmp} {\bf 71} (1987) 451.


\bibitem{Qiu}D. Gepner and Z. Qiu, {\sl \np} {\bf B285} 
[FS19] (1987) 423.






\bibitem{Lykken}  J. Lykken, {\sl \np} {\bf B313} (1989) 473.


\bibitem{Dunne} G. V. Dunne, I. G. Halliday and P. Suranyi, 
{\sl \np} {\bf B325} (1989) 526.



\bibitem{Neme} D. Nemeschansky, {\sl \np} {\bf B363} (1989) 665.


\bibitem{Bilal}A. Bilal, {\sl \pl} {\bf B226} (1989) 272.



\bibitem{Distler}J. Distler and Z. Qiu, {\sl \np} {\bf B336}
(1990) 533.



\bibitem{Ahn} C. Ahn, S. Chung and S. H. Tye,  {\sl \np} 
{\bf B365} (1991) 191.




\bibitem{Pais}A. Pais and V.
Rittenberg, {\sl \jmp} {\bf 16} (1975) 2063;  M.
Scheunert, W. Nahm and  V.
Rittenberg, {\sl \jmp} {\bf 18} (1977) 155.


\bibitem{Review} For a review on general aspects of Lie
superalgebras see M. Scheunert,  ``The Theory of Lie
Superalgebras",  {\sl Lect. Notes in Math.} 716,
(Springer-Verlag, Berlin, 1979) and  
L. Frappat, P. Sorba and A. Sciarrino, ``Dictionary on
Lie Superalgebras", hep-th/9607161.





\bibitem{ber}M. Bershadsky and H.
Ooguri, {\sl \pl} {\bf B229} (1989) 374.


\bibitem{osp} I. P. Ennes, A. V. Ramallo and J. M. S\'anchez
de Santos, {\sl \pl} {\bf B389} (1996) 485;  
{\sl \np} {\bf B491 [PM]} (1997) 574.


\bibitem{ospdos} I. P. Ennes and A. V. Ramallo, 
{\sl \np} {\bf B502 [PM]} (1997) 671.


\bibitem{rama} I. P. Ennes, P. Ramadevi,  A. V. Ramallo and 
J. M. S\'anchez de Santos, ``Duality in osp$(1\vert 2)$ Conformal
Field Theory and link invariants", hep-th/9709068 (in press in 
{\sl \ijmp}).

\bibitem{FdB} M. Fa\'a di Bruno, {\sl Quart. Jour. Pure Appl.
Math.} {\bf 1} (1857) 359.



\bibitem{Dickey} see  L. A. Dickey, {\sl ``Soliton equations and
Hamiltonian systems"}, (World Scientific, Singapore, 1991).


\bibitem{KP}V. Kac and D. Peterson, {\sl \adm} {\bf 53} (1984)
125.






\bibitem{Fateev} V. A. Fateev, {\sl \ijmp} {\bf A6} (1991) 2109.




\end{thebibliography}
\end{document}